\documentclass[aps,prd,preprint,groupedaddress,showpacs,A4paper]{revtex4-1}

\usepackage{amsmath,amssymb}
\usepackage{hyperref}
\usepackage{tikz}
\usetikzlibrary{intersections}
\usetikzlibrary{decorations.pathmorphing}
\usetikzlibrary{decorations.markings}

\tikzset{
    laser/.style={decorate, decoration={snake,segment length=5mm}, draw=black},
    electron/.style={draw=black, postaction={decorate},
        decoration={markings,mark=at position .55 with {\arrow[draw=black]{>}}}},
}

\newcommand{\de}{\delta}

\newcommand{\beq}{\begin{equation}}
\newcommand{\eeq}{\end{equation}}
\newcommand{\bea}{\begin{eqnarray}}
\newcommand{\eea}{\end{eqnarray}}
\newcommand\zn[1]{Z_2^{(#1)}}

\newcommand{\phiV}{\phi_{_{\mathrm{V}}}}
\newcommand{\aV}{a_{_{\mathrm{V}}}\!}
\newcommand{\bV}{b_{_{\mathrm{V}}}\!}
\newcommand{\adV}{a^\dag_{_{\mathrm{V}}}\!}
\newcommand{\bdV}{b^\dag_{_{\mathrm{V}}}\!}
\newcommand{\kdotx}{k{\cdot}x}
\newcommand{\pbar}{{\bar{p}}}
\newcommand{\qbar}{{\bar{q}}}

\newcommand{\intph}{\int\!\frac{{\bar{}\kern-0.45em d}^{\,3}{p}}{2E_{p}}}
\newcommand{\intp}{\int\!\frac{{\bar{}\kern-0.45em d}^{\,3}{p}}{2E^*_p}}
\newcommand{\intq}{\int\!\frac{{\bar{}\kern-0.45em d}^{\,3}{q}}{2E^*_q}}
\newcommand{\Dc}{D_{_{\mathrm{c}}}}
\newcommand{\Dl}{D_{_{\mathrm{\ell}}}}
\newcommand{\ee}{\mathrm{e}}
\newcommand{\w}[2]{w_{#1\,#2}}
\newcommand{\Jt}{\widetilde{\mathrm{J}}}
\newcommand{\Jb}{{\mathrm{J}}}

\newcommand{\bra}[1]{\langle #1|}
\newcommand{\ket}[1]{|#1\rangle}
\newcommand{\braV}[1]{{_{_{\mathrm{V}}}}\!\bra{#1}}
\newcommand{\ketV}[1]{\ket{#1}_{_{\mathrm{V}}}}

\begin{document}


\title{Propagation in an intense background}


\author{Martin~Lavelle, David~McMullan and Merfat~Raddadi}
\affiliation{School of Computing and Mathematics, University of Plymouth \\
Plymouth, PL4 8AA, UK}

\date{\today}

\begin{abstract}
The canonical quantisation of a charged particle in an intense laser background is developed and the associated  two-point function calculated. It is shown how the causal propagator can be extracted. We provide an interpretation of this propagator in terms of degenerate quantum processes. The spectral properties of the propagator are derived and we show how  multi-pole structures emerge as a perturbative feature. We also calculate the associated multiple wave-function renormalisations  to all orders in perturbation theory. Our calculations are for both circular and linearly polarised plane wave laser backgrounds.
\end{abstract}



\maketitle

\section{Introduction and motivation}
Interest in the propagation of an electron in a laser background has a long history that significantly predates the recent advances in laser technology that is opening up the field to precise experimental verification. The exact solvability for the matter fields in a plane wave background was discovered by Volkov in 1935~\cite{Volkov:1935zz}, quite soon after the birth of quantum electrodynamics (QED). The advent of the first functional laser in 1960~\cite{Maiman:1960} spurred extensive theoretical investigations into the effect of such a background on the propagation  and interactions of matter~\cite{reiss:387}. Most notably, non-linear corrections  to Compton scattering in such a background were calculated by various authors \cite{Nikishov:1964zza,Nikishov:1964zz,Brown:1964zz,Goldman1964103}. In the mid 1990s experimental demonstration  of such non-linear Compton scattering \cite{PhysRevLett.76.3116} started to emerge  which has revitalised  theoretical interest in the field, see for example \cite{Harvey:2009ry,Boca:2009zz,Heinzl:2009nd,Mackenroth:2010jr,Seipt:2010ya}. The commissioning of several petawatt facilities worldwide, such as the Vulcan 10 petawatt project \cite{Vulcan} or the 200 petawatt ambitions of ELI \cite{ELI}, adds further impetus to the theoretical community to  better understand quantum field theory in such intense backgrounds \cite{DiPiazza:2011tq}.

Given the maturity of this field it is useful here to clarify our motivation for studying these problems afresh and to identify what new perspective we hope to bring to this area. Again, we need to look back to the early days of QED. Infra-red effects in QED have a similar antiquity to the above work on laser backgrounds, starting   with the classic work of Bloch and Nordsieck~\cite{Bloch:1937pw} on soft infra-red effects followed by the work of Kinoshita \cite{Kinoshita:1962ur} and Lee and Nauenberg~\cite{Lee:1964is} on the collinear structures that arise both for massless charges and in the high intensity regime. The conclusion of these works can be summarised as follows: to arrive at a physical cross-section one must sum over all degenerate processes. This prescription  for the infra-red underpins all attempts at calculating standard model cross-sections~\cite{Brock:1993sz} but it has theoretical issues that are still hard to understand especially when there are both initial and final state degeneracies~\cite{Lavelle:2005bt}. In particular, the infra-red structures that arise from virtual corrections to the S-matrix need to cancel against a background of real degenerate (soft or collinear) processes that is indistinguishable  from the original process. The  existence of such degenerate backgrounds is, to say the least, problematic in most gauge theories, but a plane-wave  laser provides a well defined background. So, coming from this infra-red perspective, a particle propagating  through a laser is degenerate to one that emits and absorbs the same number of laser photons and thus these processes need to be summed over in order to describe the physical propagation of the charge. In this paper we want to develop this point of view and then back it up with a clear account of how such a system is quantised. Our hopes are that this will clarify some of the field theory underpinning  this area of laser physics and, in  turn,  refine the techniques needed to tackle the infra-red sector of gauge theories, see also~\cite{Dinu:2012tj}.

After this introduction we will, in section~\ref{two}, develop a perturbative route to the Volkov propagator for a scalar particle propagating in a circularly polarised laser. This will be based on diagrammatic techniques that emerge from a systematic application of QED and the concept of degeneracy. Here we will  calculate the leading and next to leading order contributions to the propagator. We will demonstrate the emergence of sideband modes within this perturbative framework already at lowest order in perturbation theory. Following that, in section~\ref{three}  we shall take a completely different approach to the propagator. We will develop the equal-time quantisation of the Volkov field. This will allow for a precise identification of states and commutators and hence an all orders description of the Volkov propagator. We will see that these all orders results agree with the perturbative calculations. This will be followed in section~\ref{four} by a brief discussion of polarisation effects by now working in a linearly polarised laser background using both of the approaches taken earlier in the paper. Following the conclusion there are two appendices where some technical results are collected.

\section{Summing degenerate processes}\label{two}

In what follows we will consider perturbative corrections to the free propagator when a charged particle is in an intense laser background. We will not consider loop corrections but rather study the effects due to the absorption and emission of laser photons. In the spirit of the aforementioned Lee-Nauenberg theorem  we will sum over all tree-level interactions with the laser which are indistinguishable from the propagator. We will not include in this paper any soft or collinear degeneracies. Such contributions would cancel against loop diagrams which we do not consider. This means that we will restrict ourselves to processes where the matter field has the same momentum, $p$, in both the initial and final states. Thus all emitted photons are necessarily  degenerate with the laser and hence having  four-momentum $k$. Furthermore,  the number of emitted photons must be equal to the number of absorbed laser photons as otherwise the matter field's initial and final momenta will not be the same.

We will focus here initially on circularly polarised laser photons described by the potential
\beq\label{cirpot}
A^\mu(x)= a_1^\mu\cos(\kdotx)+a_2^\mu\sin(\kdotx)\,,
\eeq
which satisfies the gauge condition $k{\cdot} A=0$, where the null vector $k$ will usually be taken to be pointing in the third direction: $k_\mu=k_0(1,0,0,1)$.  Furthermore, the polarisations vectors satisfy  $a_1{\cdot} a_2=0$ and  $a_1^2=a_2^2=a^2<0$. For simplicity we will consider scalar matter interacting with the laser. The interaction Lagrangian has the form
\beq\label{lintc}
{\cal L}_{\mathrm{int}}=-ieA^\mu\left(\phi^{\dag}\overleftrightarrow{\partial_\mu} \phi \right)+e^2A^2\phi^{\dag}\phi\,.
\eeq
To build up a perturbative description of the interaction with the laser background we use the Feynman diagram technique whereby we identify the basic interaction terms and their corresponding Feynman rules. For scalar matter two basic interactions arise corresponding to the two terms in the Lagrangian~(\ref{lintc}). The cubic term corresponds to the expected three point interaction while the quartic term generates a four point vertex which is commonly called a seagull term.

The four point vertex Feynman rule is obtained from the Fourier transform of the time-ordered product of free fields with the insertion of the four point interaction Lagrangian vertex
\beq\label{4ptv}
\int d^4x\, e^{-ip{\cdot} x}\left<0\right|T \phi(x)\phi^{\dag}(0)ie^2\int d^4z a^2\phi^{\dag}(z)\phi(z)\left|0\right>
\eeq
which after standard contractions generates
\beq\label{97121340}
ie^2a^2\left(\frac i{p^2-m^2+i\epsilon}\right)^2\,.
\eeq
The $i\epsilon$ term in this expression arises through the use of time-ordering and encapsulates  the causal nature of the propagator. At a more basic level, the significance of time ordering in (\ref{4ptv}) reflects basic commutators of the fields, which are well understood for this scalar field.

The Feynman rule is then obtained from this by amputating the matter lines  resulting in Fig.~1. Note that it is a special property of circularly polarised lasers, where $A^2=a^2$, that the only seagull term which survives is when one of the laser lines corresponds to absorbtion and one to emission and so both matter lines in (\ref{97121340}) have momentum $p$. This is not the case for, e.g., linearly polarised lasers as we will see later in section~\ref{four}.

\begin{figure}
\begin{minipage}[c]{0.2\linewidth}
\begin{tikzpicture}[xscale=0.6, yscale=0.75]
 \draw[electron] (0,0) -- node[below=4.5pt]{$p$}(2,0);
 \draw[laser] (0,1.5) -- (2,0);
 \draw[laser] (4,1.5)-- (2,0);
 \draw[electron] (2,0) -- node[below=4.5pt]{$p$}(4,0);
 \fill (2, 0) circle (1.5pt);
 \draw[white] (4,0) -- node [right=2pt]
{$\displaystyle\color{black}=
ie^2a^2$}(4,1.5);
\end{tikzpicture}
\end{minipage}
\caption{\label{seagullcircrule} The seagull interaction}
\end{figure}

In a similar way, through insertion of the three point interaction Lagrangian vertex, one obtains the Feynman rules corresponding to Fig.'s~2a and~b.

\begin{figure}
\begin{minipage}[c]{0.35\linewidth}
\begin{tikzpicture}[xscale=0.6, yscale=0.75]
 \draw[electron] (0,0) -- node[below=4.5pt]{$p$}(2,0);
 \draw[laser] (0,1.5) -- (2,0);
 \draw[electron] (2,0) -- node[below=2pt]{$p+k$}(4,0);
 \fill (2, 0) circle (1.5pt);
 \draw[white] (4,0) -- node [right=2pt]
{$\displaystyle\color{black}=
ie(a_1+ia_2){\cdot} p$}(4,1.5);
\end{tikzpicture}
\centerline{(a)}
\end{minipage}
\quad
\begin{minipage}[c]{0.35\linewidth}
 \begin{tikzpicture}[xscale=0.6, yscale=0.75]
 \draw[electron] (0,0) -- node[below=4.5pt]{$p$}(2,0);
 \draw[laser] (4,1.5) -- (2,0);
 \draw[electron] (2,0) -- node[below=2pt]{$p-k$}(4,0);
 \fill (2, 0) circle (1.5pt);
 \draw[white] (4,0) -- node [right=2pt]
{$\displaystyle\color{black}=
ie(a_1-ia_2){\cdot} p$}(4,1.5);
\end{tikzpicture}
\newline
\centerline{(b)}
\end{minipage}
\caption{\label{threecircrule} The three point interactions}
\end{figure}

We should here clarify the diagrammatic convention which we follow: laser lines coming from the left correspond to absorbtion of laser photons while lines going out to the right denote emission processes where the emitted photons are degenerate with the laser.

The leading order corrections to the propagator are given by the seagull term of Fig.~1 and the diagrams in Fig.~3. We note that should the initial and final matter field have a different momentum, see, e.g., the diagrams of Fig.~4, then these contributions can appear in odd powers of the coupling $e$ and it is clear that adding these processes to those of Fig.~3 would yield a result non-analytic in $\alpha$.
These are, anyhow,  non-degenerate with the processes of Fig.~3 and cannot contribute to the propagator.

In this spirit we now consider the leading and next to leading perturbative corrections to the free propagator with the matter field having the same initial and final momentum.

%
%
%
%
%
%
%
%
%
%

\bigskip

\subsection{The propagator at order $e^2$}

\begin{figure}
\begin{minipage}[c]{0.25\linewidth}
\begin{tikzpicture}[xscale=0.6, yscale=0.75]
 \draw[electron] (0,0) -- node[below=4.5pt]{$p$}(2,0);
 \draw[laser] (0,1.5) -- (2,0);
 \draw[laser](6,1.5) -- (4,0);
 \draw[electron] (2,0) -- node[below=2pt]{$p+k$}(4,0);
 \draw[electron] (4,0) -- node[below=4.5pt]{$p$}(6,0);
 \fill (2, 0) circle (1.5pt);
 \fill (4,0) circle (1.5pt);
\end{tikzpicture}
\newline
\centerline{(a)}
\end{minipage}
\quad
\begin{minipage}[c]{0.25\linewidth}
\begin{tikzpicture}[xscale=0.6, yscale=0.75]
 \draw[electron] (0,0) -- node[below=4.5pt]{$p$}(2,0);
 \draw[laser](4,1.5) -- (2,0);
 \fill[white] (2.95,0.78) circle (2.5pt);
\draw[laser] (2,1.5) -- (4,0);
 \draw[electron] (2,0) -- node[below=2pt]{$p-k$}(4,0);
 \draw[electron] (4,0) -- node[below=4.5pt]{$p$}(6,0);
 \fill (2, 0) circle (1.5pt);
 \fill (4,0) circle (1.5pt);
\end{tikzpicture}
\newline
\centerline{(b)}
\end{minipage}
\caption{\label{line2propdiags} Leading order contributions to the propagator}
\end{figure}

\begin{figure}
\begin{minipage}[c]{0.25\linewidth}
  \begin{tikzpicture}[xscale=0.6, yscale=0.75]
 \draw[electron] (0,0) -- node[below=4.5pt]{$p$}(2,0);
 \draw[laser] (0,1.5) -- (2,0);
 \draw[electron] (2,0) -- node[below=2pt]{$p+k$}(4,0);
 \fill (2, 0) circle (1.5pt);
\end{tikzpicture}\newline
\centerline{(a)}
\end{minipage}
\quad
\begin{minipage}[c]{0.25\linewidth}
\begin{tikzpicture}[xscale=0.6, yscale=0.75]
 \draw[electron] (0,0) -- node[below=4.5pt]{$p$}(2,0);
 \draw[laser] (0,1.5) -- (2,0);
 \draw[laser] (4,1.5) -- (2,0);
 \draw[laser](6,1.5) -- (4,0);
 \draw[electron] (2,0) -- node[below=4.5pt]{$p$}(4,0);
 \draw[electron] (4,0) -- node[below=2pt]{$p+k$}(6,0);
 \fill (2, 0) circle (1.5pt);
 \fill (4,0) circle (1.5pt);
\end{tikzpicture}
\newline
\centerline{(b)}
\end{minipage}
\quad
\begin{minipage}[c]{0.25\linewidth}
\begin{tikzpicture}[xscale=0.6, yscale=0.75]
 \draw[electron] (0,0) -- node[below=4.5pt]{$p$}(2,0);
 \draw[laser](0,1.5) -- (2,0);
\draw[laser] (2,1.5) -- (4,0);
\draw[laser] (8,1.5) -- (6,0);
 \draw[electron] (2,0) -- node[below=2pt]{$p+k$}(4,0);
 \draw[electron] (4,0) -- node[below=2pt]{$p+2k$}(6,0);
 \draw[electron] (6,0) -- node[below=2pt]{$p+k$}(8,0);
 \fill (2, 0) circle (1.5pt);
 \fill (4,0) circle (1.5pt);
 \fill (6,0) circle (1.5pt);
\end{tikzpicture}
\newline
\centerline{(c)}
\end{minipage}
\caption{Some leading and next to leading non-degenerate laser-particle interactions}
\end{figure}
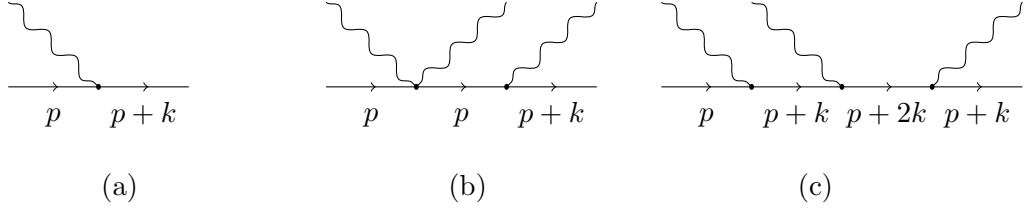

The leading order corrections to the free propagator are given by Fig.~1 and the two diagrams of Fig.~3.
Fig.~1, with the four-point vertex, contributes
\beq\label{dp1}
\frac{i}{p^2-m^2}i e^2 a^2 \frac{i}{p^2-m^2}\,.
\eeq
Note that in this, and for the rest of this section, we have suppressed the $i\epsilon$ term for compactness of presentation.
Expression (\ref{dp1}) is   a double pole in $p^2-m^2$ and this corresponds to a mass shift effect. Writing the propagator as the free propagator plus tree diagram corrections \beq\label{eqn4712}
\frac{i}{p^2-m^2-\delta m^2}=\frac{i}{p^2-m^2}\left[1+ \frac{i}{p^2-m^2}(-i\delta m^2)+\dots\right]\,,
\eeq
we can immediately identify the mass shift as
\beq\label{eqn2712}
\delta m^2=-e^2 a^2\,,
\eeq
which is a positive correction as $a^2$ is negative. This is the mass shift that was first identified by Sengupta \cite{Sengupta1952} and later rediscovered  by various authors~\cite{Nikishov:1964zz,Kibble:1965zza}, that arises when a charge is in such a laser background resulting in an effective mass $m_*$ where
\begin{equation}\label{mstar}
    m_*^2=m^2+e^2(-a^2)\,.
\end{equation}

The matter lines between the absorbtion and emission vertices in the  two diagrams in Fig.~3 carry momentum $p\pm k$. Putting the diagrams together we have
\beq\label{square}
-e^2\left(\frac i{p^2-m^2}\right)^2\left(\left(a_1{\cdot} p\right)^2+\left(a_2{\cdot} p\right)^2\right)
\left(\frac i{(p+k)^2-m^2}+\frac i{(p-k)^2-m^2}\right)\,.
\eeq
For convenience we define
\beq
w_p^2=\frac{\left(a_1{\cdot} p\right)^2+\left(a_2{\cdot} p\right)^2}{(p{\cdot} k)^2}\,,
\eeq
and note that when the terms in the large brackets in (\ref{square}) are placed on a common denominator the resulting numerator cancels one of the poles in $p^2-m^2$ yielding the contribution
\beq
2ie^2 w_p^2 \frac{(p{\cdot} k)^2}{\left(p^2-m^2\right)\left((p+k)^2-m^2\right)\left((p-k)^2-m^2\right)}\,.
\eeq
At this stage we use partial fractions to disentangle the  fraction above:
\begin{align*}
    \frac{(p{{\cdot}} k)^2}{\left(p^2-m^2\right)\left((p+k)^2-m^2\right)\left((p-k)^2-m^2\right)}=&-\frac 1{4(p^2-m^2)}\\&+
\frac 1{8((p+k)^2-m^2)}+\frac 1{8((p-k)^2-m^2)}\,.
\end{align*}

We identify three structures here. The first is a single pole in $p^2-m^2$  and thus a wave-function renormalisation effect on the initial pole. The other two terms are new poles at $(p\pm k)^2-m^2$. These poles first appear at order $e^2$ and are of exactly the form obtained in \cite{Reiss:1966A,Eberly:1966b} which were obtained using non-perturbative arguments. We emphasise that they are completely perturbative and that they appear here from the lowest order correction to the free propagator. We will see in a moment that they also acquire the mass shift of (\ref{eqn2712}) from diagrams entering at order $e^4$. We will write the terms involving poles in $p^2-m^2$ in the form
\beq\label{eqnHRY2343}
\frac {i \zn0}{p^2-m^2-\de m^2}=\frac{i\zn0}{p^2-m^2}\left[
1+\frac i{p^2-m^2}\left(-i\de m^2\right)+\dots\right] \,,
\eeq
where the bracketed superscript in $\zn0$ indicates that this corresponds to a denominator  $p^2-m^2=(p+0k)^2-m^2$. From this equation we see that at this order the residue of the pole is
\beq
\zn0(w_p) =1-e^2\frac{w_p^2}2\,,
\eeq
which is a finite renormalisation.

The new pole structures at $(p\pm nk)^2-m^2$ first appear, for integer $n$, at order $e^{2n}$ and  the appropriate wave function renormalisation type constant, $\zn{\pm n}$, will be a power series in the coupling starting at order $e^{2n}$. From the above we have at leading order
\beq
\zn{\pm1}(w_p)=e^2\frac{w_p^2}4\,.
\eeq
There is an important point to note here that we will return to later in the paper. In QED without a background the wave-function renormalisation constant $Z_2$ is momentum independent. This is because it is identified with the residue of the appropriate pole, i.e, it is evaluated on-shell. Relativistic invariance then dictates that it could only be a function of $p^2$ which becomes $m^2$ in $Z_2$ irrespective of the choice of positive or negative energy pole. With a background things are different as we have an additional vector $k$ that can, and indeed does,  enter into the calculations. This means that the wave-function renormalisations are now not constants but can be functions of $p{\cdot}k$, $p{\cdot}a_1$ and $p{\cdot}a_2$, with $p$ on-shell. In addition, the residue will be different depending on whether the positive or negative energy pole is picked up.

\subsection{The propagator at order $e^4$}

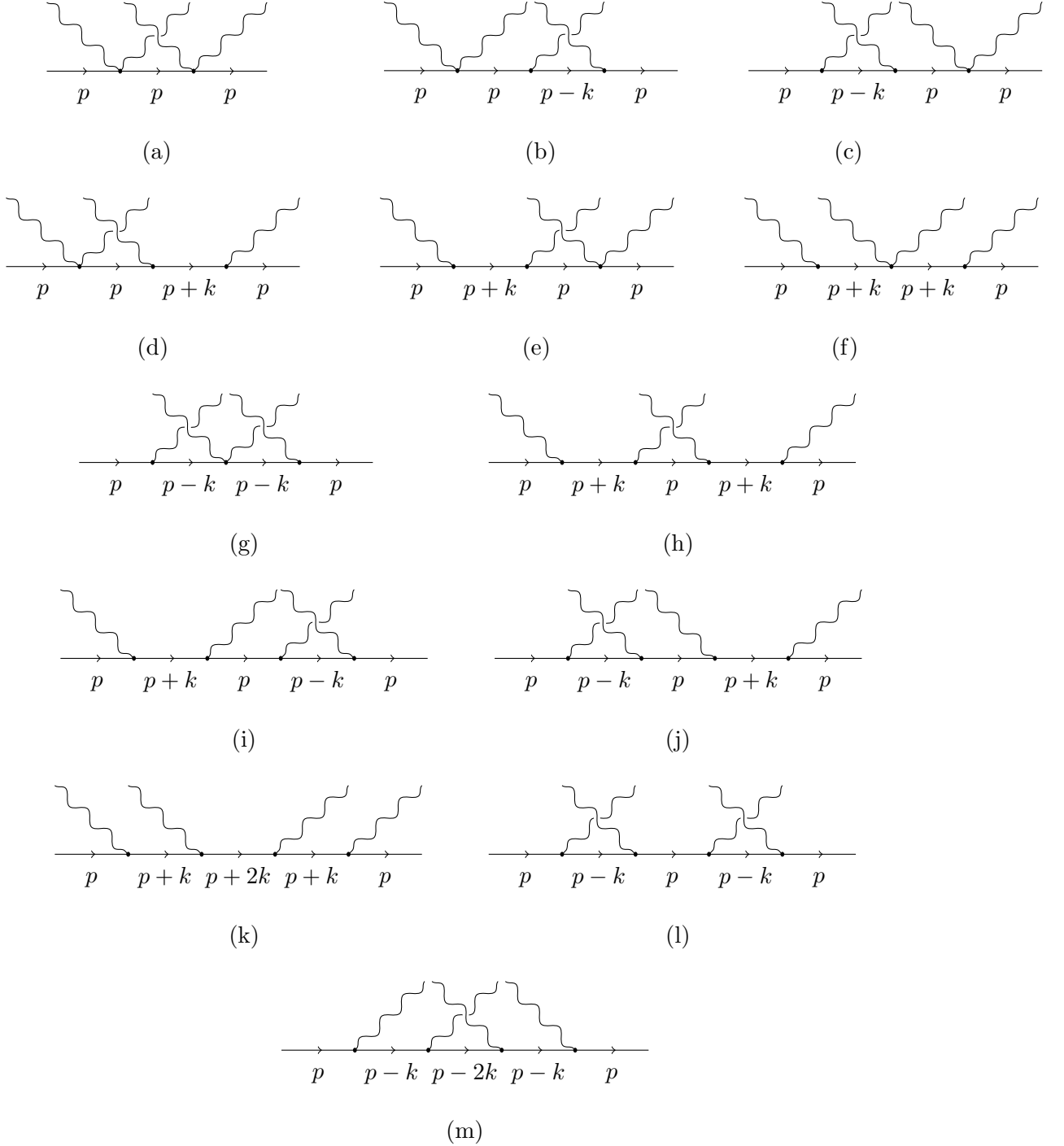
\begin{figure}
\begin{minipage}[c]{0.35\linewidth}
\begin{tikzpicture}[xscale=0.6, yscale=0.75]
 \draw[electron] (0,0) -- node[below=4.5pt]{$p$}(2,0);
 \draw[electron] (2,0) -- node[below=4.5pt]{$p$}(4,0);
 \draw[electron] (4,0) -- node[below=4.5pt]{$p$}(6,0);
 \draw[laser] (0,1.5) -- (2,0);
 \draw[laser](3.9,1.5) -- (2,0);
 \fill[white] (3.02,0.78) circle (2.5pt);
 \draw[laser] (2.1,1.5) -- (4,0);
 \draw[laser](6,1.5) -- (4,0);
 \fill (2, 0) circle (1.5pt);
 \fill (4,0) circle (1.5pt);
\end{tikzpicture}
\centerline{(a)}
\end{minipage}
\quad
\begin{minipage}[c]{0.35\linewidth}
\begin{tikzpicture}[xscale=0.6, yscale=0.75]
\draw[electron] (0,0) -- node[below=4.5pt]{$p$}(2,0);
 \draw[electron] (2,0) -- node[below=4.5pt]{$p$}(4,0);
 \draw[electron] (4,0) -- node[below=2pt]{$p-k$}(6,0);
 \draw[electron] (6,0) -- node[below=4.5pt]{$p$}(8,0);
 \draw[laser] (0,1.5) -- (2,0);
 \draw[laser](3.9,1.5) -- (2,0);
 \draw[laser] (6,1.5) -- (4,0);
  \fill[white] (5.02,0.78) circle (2.5pt);
 \draw[laser](4.1,1.5) -- (6,0);
 \fill (2, 0) circle (1.5pt);
 \fill (4,0) circle (1.5pt);
 \fill (6, 0) circle (1.5pt);
\end{tikzpicture}
\newline
\centerline{(b)}
\end{minipage}
\quad
\begin{minipage}[c]{0.2\linewidth}
\begin{tikzpicture}[xscale=0.6, yscale=0.75]
\draw[electron] (0,0) -- node[below=4.5pt]{$p$}(2,0);
 \draw[electron] (2,0) -- node[below=2pt]{$p-k$}(4,0);
 \draw[electron] (4,0) -- node[below=4.5pt]{$p$}(6,0);
 \draw[electron] (6,0) -- node[below=4.5pt]{$p$}(8,0);
 \draw[laser] (4.1,1.5) -- (6,0);
 \draw[laser](8,1.5) -- (6,0);
 \draw[laser] (3.9,1.5) -- (2,0);
  \fill[white] (2.97,0.78) circle (2.5pt);
 \draw[laser](2,1.5) -- (4,0);
 \fill (2, 0) circle (1.5pt);
 \fill (4,0) circle (1.5pt);
 \fill (6, 0) circle (1.5pt);
\end{tikzpicture}
\newline
\centerline{(c)}
\end{minipage}
\vskip5mm
\begin{minipage}[c]{0.35\linewidth}
\begin{tikzpicture}[xscale=0.6, yscale=0.75]
\draw[electron] (0,0) -- node[below=4.5pt]{$p$}(2,0);
 \draw[electron] (2,0) -- node[below=4.5pt]{$p$}(4,0);
 \draw[electron] (4,0) -- node[below=2pt]{$p+k$}(6,0);
 \draw[electron] (6,0) -- node[below=4.5pt]{$p$}(8,0);
 \draw[laser] (0,1.5) -- (2,0);
 \draw[laser](3.9,1.5) -- (2,0);
  \fill[white] (3,0.78) circle (2.5pt);
 \draw[laser] (8,1.5) -- (6,0);
 \draw[laser](2.1,1.5) -- (4,0);
 \fill (2, 0) circle (1.5pt);
 \fill (6,0) circle (1.5pt);
 \fill (4, 0) circle (1.5pt);
\end{tikzpicture}
\centerline{(d)}
\end{minipage}
\quad
\begin{minipage}[c]{0.35\linewidth}
\begin{tikzpicture}[xscale=0.6, yscale=0.75]
\draw[electron] (0,0) -- node[below=4.5pt]{$p$}(2,0);
 \draw[electron] (2,0) -- node[below=2pt]{$p+k$}(4,0);
 \draw[electron] (4,0) -- node[below=4.5pt]{$p$}(6,0);
 \draw[electron] (6,0) -- node[below=4.5pt]{$p$}(8,0);
 \draw[laser] (0,1.5) -- (2,0);
 \draw[laser](6,1.5) -- (4,0);
  \fill[white] (4.97,0.78) circle (2.5pt);
 \draw[laser] (4,1.5) -- (6,0);
 \draw[laser](8,1.5) -- (6,0);
 \fill (2, 0) circle (1.5pt);
 \fill (6,0) circle (1.5pt);
 \fill (4, 0) circle (1.5pt);
\end{tikzpicture}
\newline
\centerline{(e)}
\end{minipage}
\quad
\begin{minipage}[c]{0.2\linewidth}
\begin{tikzpicture}[xscale=0.6, yscale=0.75]
\draw[electron] (0,0) -- node[below=4.5pt]{$p$}(2,0);
 \draw[electron] (2,0) -- node[below=2pt]{$p+k$}(4,0);
 \draw[electron] (4,0) -- node[below=2pt]{$p+k$}(6,0);
 \draw[electron] (6,0) -- node[below=4.5pt]{$p$}(8,0);
 \draw[laser] (0,1.5) -- (2,0);
 \draw[laser](2,1.5) -- (4,0);
 \draw[laser] (6,1.5) -- (4,0);
 \draw[laser](8,1.5) -- (6,0);
 \fill (2, 0) circle (1.5pt);
 \fill (4,0) circle (1.5pt);
 \fill (6, 0) circle (1.5pt);
\end{tikzpicture}
\newline
\centerline{(f)}
\end{minipage}
\quad
\vskip5mm
\begin{minipage}[c]{0.4\linewidth}
\begin{tikzpicture}[xscale=0.6, yscale=0.75]
\draw[electron] (0,0) -- node[below=4.5pt]{$p$}(2,0);
 \draw[electron] (2,0) -- node[below=2pt]{$p-k$}(4,0);
 \draw[electron] (4,0) -- node[below=2pt]{$p-k$}(6,0);
 \draw[electron] (6,0) -- node[below=4.5pt]{$p$}(8,0);
 \draw[laser] (3.9,1.5) -- (2,0);
 \fill[white] (2.96,0.78) circle (2.5pt);
 \draw[laser](2,1.5) -- (4,0);
 \draw[laser](6,1.5) -- (4,0);
\fill[white] (5.02,0.78) circle (2.5pt);
\draw[laser] (4.1,1.5) -- (6,0);
 \fill (2, 0) circle (1.5pt);
 \fill (4,0) circle (1.5pt);
 \fill (6, 0) circle (1.5pt);
\end{tikzpicture}
\newline
\centerline{(g)}
\end{minipage}
\quad
\begin{minipage}[c]{0.4\linewidth}
\begin{tikzpicture}[xscale=0.6, yscale=0.75]
 \draw[electron] (0,0) -- node[below=4.5pt]{$p$}(2,0);
 \draw[electron] (2,0) -- node[below=2pt]{$p+k$}(4,0);
 \draw[electron] (4,0) -- node[below=4.5pt]{$p$}(6,0);
 \draw[electron] (6,0) -- node[below=2pt]{$p+k$}(8,0);
 \draw[electron] (8,0) -- node[below=4.5pt]{$p$}(10,0);
 \draw[laser] (0,1.5) -- (2,0);
 \draw[laser](5.9,1.5) -- (4,0);
 \fill[white] (5.02,0.78) circle (2.5pt);
 \draw[laser] (4.1,1.5) -- (6,0);
 \draw[laser](10,1.5) -- (8,0);
 \fill (2, 0) circle (1.5pt);
 \fill (4,0) circle (1.5pt);
 \fill (6, 0) circle (1.5pt);
 \fill (8,0) circle (1.5pt);
\end{tikzpicture}
\newline
\centerline{(h)}
\end{minipage}
\qquad \vskip5mm
\begin{minipage}[c]{0.4\linewidth}
\begin{tikzpicture}[xscale=0.6, yscale=0.75]
\draw[electron] (0,0) -- node[below=4.5pt]{$p$}(2,0);
 \draw[electron] (2,0) -- node[below=2pt]{$p+k$}(4,0);
 \draw[electron] (4,0) -- node[below=4.5pt]{$p$}(6,0);
 \draw[electron] (6,0) -- node[below=2pt]{$p-k$}(8,0);
 \draw[electron] (8,0) -- node[below=4.5pt]{$p$}(10,0);
 \draw[laser] (0,1.5) -- (2,0);
 \draw[laser](5.9,1.5) -- (4,0);
 \draw[laser] (8,1.5) -- (6,0);
  \fill[white] (6.95,0.78) circle (2.5pt);
 \draw[laser](6,1.5) -- (8,0);
 \fill (2, 0) circle (1.5pt);
 \fill (4,0) circle (1.5pt);
 \fill (6, 0) circle (1.5pt);
 \fill (8,0) circle (1.5pt);
\end{tikzpicture}
\centerline{(i)}
\end{minipage}
\quad
\begin{minipage}[c]{0.4\linewidth}
\begin{tikzpicture}[xscale=0.6, yscale=0.75]
\draw[electron] (0,0) -- node[below=4.5pt]{$p$}(2,0);
 \draw[electron] (2,0) -- node[below=2pt]{$p-k$}(4,0);
 \draw[electron] (4,0) -- node[below=4.5pt]{$p$}(6,0);
 \draw[electron] (6,0) -- node[below=2pt]{$p+k$}(8,0);
 \draw[electron] (8,0) -- node[below=4.5pt]{$p$}(10,0);
 \draw[laser] (3.9,1.5) -- (2,0);
 \fill[white] (2.95,0.78) circle (2.5pt);
 \draw[laser](2,1.5) -- (4,0);
 \draw[laser] (4.1,1.5) -- (6,0);
 \draw[laser](10,1.5) -- (8,0);
 \fill (2, 0) circle (1.5pt);
 \fill (4,0) circle (1.5pt);
 \fill (6, 0) circle (1.5pt);
 \fill (8,0) circle (1.5pt);
\end{tikzpicture}
\centerline{(j)}
\end{minipage}
\quad \vskip5mm
\begin{minipage}[c]{0.4\linewidth}
\begin{tikzpicture}[xscale=0.6, yscale=0.75]
 \draw[electron] (0,0) -- node[below=4.5pt]{$p$}(2,0);
 \draw[electron] (2,0) -- node[below=2pt]{$p+k$}(4,0);
 \draw[electron] (4,0) -- node[below=2pt]{$p+2k$}(6,0);
 \draw[electron] (6,0) -- node[below=2pt]{$p+k$}(8,0);
 \draw[electron] (8,0) -- node[below=4.5pt]{$p$}(10,0);
 \draw[laser] (0,1.5) -- (2,0);
 \draw[laser](2,1.5) -- (4,0);
 \draw[laser] (8,1.5) -- (6,0);
 \draw[laser](10,1.5) -- (8,0);
 \fill (2, 0) circle (1.5pt);
 \fill (4,0) circle (1.5pt);
 \fill (6, 0) circle (1.5pt);
 \fill (8,0) circle (1.5pt);
\end{tikzpicture}
\newline
\centerline{(k)}
\end{minipage}
\quad
\begin{minipage}[c]{0.4\linewidth}
\begin{tikzpicture}[xscale=0.6, yscale=0.75]
 \draw[electron] (0,0) -- node[below=4.5pt]{$p$}(2,0);
 \draw[electron] (2,0) -- node[below=2pt]{$p-k$}(4,0);
 \draw[electron] (4,0) -- node[below=4.5pt]{$p$}(6,0);
 \draw[electron] (6,0) -- node[below=2pt]{$p-k$}(8,0);
 \draw[electron] (8,0) -- node[below=4.5pt]{$p$}(10,0);
 \draw[laser] (4,1.5) -- (2,0);
  \fill[white] (2.95,0.78) circle (2.5pt);
 \draw[laser](2,1.5) -- (4,0);
 \draw[laser] (8,1.5) -- (6,0);
  \fill[white] (6.95,0.78) circle (2.5pt);
 \draw[laser](6,1.5) -- (8,0);
 \fill (2, 0) circle (1.5pt);
 \fill (4,0) circle (1.5pt);
 \fill (6, 0) circle (1.5pt);
 \fill (8,0) circle (1.5pt);
\end{tikzpicture}
\newline
\centerline{(l)}
\end{minipage}
\quad
\vskip5mm
\begin{minipage}[c]{0.4\linewidth}
\begin{tikzpicture}[xscale=0.6, yscale=0.75]
 \draw[electron] (0,0) -- node[below=4.5pt]{$p$}(2,0);
 \draw[electron] (2,0) -- node[below=2pt]{$p-k$}(4,0);
 \draw[electron] (4,0) -- node[below=2pt]{$p-2k$}(6,0);
 \draw[electron] (6,0) -- node[below=2pt]{$p-k$}(8,0);
 \draw[electron] (8,0) -- node[below=4.5pt]{$p$}(10,0);
 \draw[laser] (3.9,1.5) -- (2,0);
 \draw[laser](5.9,1.5) -- (4,0);
 \fill[white] (5.02,0.78) circle (2.5pt);
 \draw[laser] (4.1,1.5) -- (6,0);
 \draw[laser](6.1,1.5) -- (8,0);
 \fill (2, 0) circle (1.5pt);
 \fill (4,0) circle (1.5pt);
 \fill (6, 0) circle (1.5pt);
 \fill (8,0) circle (1.5pt);
\end{tikzpicture}
\centerline{(m)}
\end{minipage}
\caption{\label{nnl}The ${\cal O}(e^4)$ degenerate contributions to the propagator}
\end{figure}

We now proceed to the next order correction. These are given by the thirteen diagrams of Fig.~\ref{nnl}.

We note from these diagrams that the internal matter lines may carry momentum $p$, $p\pm k$ and, now, $p\pm 2k$. Thus we can expect the poles from the lowest order diagrams to receive corrections and additionally new poles at $(p\pm 2k)^2-m^2$ to arise. It is helpful to divide up the study of these diagrams according to the number of four point and of three point vertices in the diagrams. Diagram Fig.~\ref{nnl}a which only contains the four point vertex only features poles in $p^2-m^2$. It  simply generates the next order term in the geometric series of the mass shift found above in~(\ref{eqn4712}).  In other words~(\ref{eqn2712}) is not altered by a higher order correction:
\beq
\frac{i}{p^2-m^2-\delta m^2}=\frac{i}{p^2-m^2}\left[1+ \frac i{p^2-m^2}(-i\delta m^2)+
\left(\frac i{p^2-m^2}\right)^2(-i\delta m^2)^2+
\dots\right]\,.
\eeq

The next diagrams, Fig.~\ref{nnl}b-g, with a mix of four point and three point vertices only generate double poles (some single poles cancel in the sum of these diagrams). They yield
\beq\label{Hry0000}
e^4\frac{a^2 w_p^2}{p{\cdot} k}
\left[
-i\frac{p{\cdot} k}2\left(\frac i{p^2-m^2}\right)^2
+i\frac{p{\cdot} k}4\left(\frac i{(p+k)^2-m^2}\right)^2
+i\frac{p{\cdot} k}4\left(\frac i{(p-k)^2-m^2}\right)^2
\right]\,.
\eeq
Finally at this order, the remaining diagrams of Fig.~\ref{nnl}, without four point vertices, only generate single pole structures (with double pole terms cancelling in the sum of these diagrams). They yield
\begin{align*}
e^4w_p^4\Big[
\frac3 {32} \frac i{p^2-m^2}-\frac1{16}\frac i{(p+k)^2-m^2}-&\frac1{16}\frac i{(p-k)^2-m^2}
\\ &+\frac1{64}\frac i{(p+2k)^2-m^2}+\frac1{64}\frac i{(p-2k)^2-m^2}
\Big]
\,.
\end{align*}

These last two equations clearly display the behaviour stated in Reiss \cite{Reiss:1966A} that the four point vertices solely generate the mass shift while the three point vertices yield wave function renormalisation structures. We will also see how this behaviour arises in the next section from a canonical quantisation approach. However, later in section~\ref{four}, we will see that this divide does not hold for linearly polarised lasers.

The cross terms in (\ref{eqnHRY2343}) involving the order $e^2$ mass shift and $\zn{\pm1}$ can be recognised in (\ref{Hry0000}) showing again that the mass shift is determined by the lowest order terms and receives no additional corrections. In the next section we will understand at all orders why these new poles acquire the same  mass shift (\ref{eqn2712}).  The remaining terms above tell us that
\beq
\zn0(w_p)=1-e^2\frac{w_p^2}2+ e^4\frac{3w_p^4}{32}\,,
\eeq
that the next correction to the $(p\pm k)^2-m^2$ wave function renormalisation is
\beq
\zn{\pm1}(w_p)=e^2\frac{ w_p^2}4-e^4\frac{w_p^4}{16}\,,
\eeq
and that the first term of the wave function renormalisation $(p\pm 2k)^2-m^2$ is
\beq
\zn{\pm2}(w_p)=e^4\frac{w_p^4}{64}\,.
\eeq
We will proceed to obtain a better understanding of all of these structures in the next section when we investigate the two-point function for the Volkov field.

\section{Time ordered products and the Volkov propagator }\label{three}

In this section we want to analyse in a more fundamental way the Volkov propagator and, in so doing, we will derive the all-orders generalisation of the perturbative results obtained in the previous section using the Feynman diagram method. Again, we consider a circularly polarised (plane-wave) laser described by the classical potential (\ref{cirpot})
and identify the (complex) Volkov quantum field $\phiV(x)$ as a solution to the background laser coupled Klein-Gordon equation:
\begin{equation}\label{KGA}
    (D^2+m^2)\phiV(x)=0\,,
\end{equation}
with $D^2=D_\mu D^\mu$ and our conventions are such that $D_\mu=\partial_\mu+ieA_\mu$.

In order to understand how (\ref{KGA}) can be solved it is useful to recall the method used  in the free case where we wish to solve the Klein-Gordon equation $(\partial^2+m^2)\phi=0$ with now $\partial^2=\partial_\mu\partial^\mu$. The solution to this is built up by noting that the plane waves $\ee^{\pm ip{\cdot}x}$ satisfy the equation $\partial^2\ee^{\pm ip{\cdot}x}=-p^2\ee^{\pm ip{\cdot}x}$. So, if the four momentum $p$ is taken to be on-shell: $p\to\hat{p}=(E_p,\underline{p})$ where $E_p=\sqrt{|\underline{p}|^2+m^2}$ so that $\hat{p}^2=m^2$, then a superposition of these plane-waves gives us the general solution to the Klein-Gordon equation
\begin{equation}\label{KG}
    \phi(x)=\intph\left(\ee^{-i\hat{p}{\cdot}x}a(p)+\ee^{i\hat{p}{\cdot}x}b^\dag(p)\right)\,,
\end{equation}
where we have  adopted the convenient notation that ${\ \bar{}\kern-0.45em d}^{\,3}{p}=d^3p/(2\pi)^3$. This is a quantum field because the modes $a(p)$ and $b^\dag(p)$ are quantum operators whose commutator properties are central to the construction of the corresponding states and their propagation.

In the Volkov case a similar procedure  is possible. That is, we can replace, for example, $\ee^{-ip{\cdot x}}$ by a function $E(x,p)$ where $D^2E(x,p)=-p^2 E(x,p)$ (see, for example, \cite{Ritus:1972ky}). So again, if the momentum is taken on-shell as above, then we can build up a solution to (\ref{KGA}) using $E(x,\hat{p})$. But, in the light of our discussions in the previous section, we expect that the  on-shell condition in such a laser background should involve the effective mass~(\ref{mstar}). This suggests that we should really be looking for a replacement of the plane wave by a distorted field $D(x,p)$ such that now
\begin{equation}\label{KGAD}
    D^2D(x,p)=(-p^2+\delta m^2)D(x,p)\,.
\end{equation}
Then if we go on-shell at the starred-mass: $p\to \pbar$ with now $\pbar=(E^*_p,\underline{p})$ where $E^*_p=\sqrt{|\underline{p}|^2+m_*^{2}}$ so that $\pbar^2=m_*^{2}=m^2+\delta m^2$, we will still satisfy the coupled Klein-Gordon equation with mass $m$:
\begin{equation}\label{disteqn}
    D^2D(x,\pbar)=(-\pbar^2+\delta m^2)D(x,\pbar)=-m^2D(x,\pbar)\,,
\end{equation}
and in this way build up the Volkov field. Obviously these two approaches are not unrelated and indeed we have
\begin{equation}\label{eandd}
    E(x,p)=\exp\left({-i\frac{\delta m^2}{2p{\cdot}k}\kdotx}\right)D(x,p)\,.
\end{equation}
But, as we shall see, the transition from on-shell to off-shell momentum is a subtle point in our analysis and focussing  on the distortion field $D(x,\pbar)$ will make this step much clearer.

Specialising now to the circular polarisation case, where we write the distortion field as $\Dc(x,\pbar)$, we find that
\begin{equation}\label{dist}
    \Dc(x,\pbar)=\ee^{-i\pbar{\cdot}x}\ee^{ie(\w{1}{\pbar}\sin(\kdotx)+\w{2}{\pbar}\cos(\kdotx))}
\end{equation}
where
\begin{equation}\label{w1and2}
    \w{1}{\pbar}=-\frac{\pbar{\cdot}a_1}{\pbar{\cdot}k}\qquad \mathrm{and} \qquad \w{2}{\pbar}=\frac{\pbar{\cdot}a_2}{\pbar{\cdot}k}\,.
\end{equation}
Using these, the equation of motion (\ref{KGA}) can then be exactly solved  and we find that the Volkov field can be written as
\begin{equation}\label{phivol}
    \phiV(x)=\intp\left(\Dc(x,\pbar)\aV(p)+\Dc(x,-\pbar)\bdV(p)\right)\,.
\end{equation}

The momentum assignment in (\ref{phivol}) might seem a little unnatural as we could equally well satisfy the equation of motion (\ref{KGA}) by using $\Dc(x,\pbar)^\dag\bdV(p)$ in the second term. But, as we shall see, this would then obstruct the  gauge transformation properties of the Volkov field.

To understand this it is useful to note that there is a residual gauge freedom in our description of the background vector potential $A_\mu$ and we could equally well use $A_\mu(x)+\partial_\mu\theta(x)$ where
\begin{equation}\label{gt}
    \theta(x)=\theta_1\sin(\kdotx)+\theta_2\cos(\kdotx)\,,
\end{equation}
with $\theta_1$ and $\theta_2$ arbitrary constants. This gauge transformation can equivalently be written in terms of the polarisation vectors as
\begin{equation}\label{gtpol}
    a_1^\mu\to a_1^\mu+\theta_1k^\mu\qquad \mathrm{and} \qquad a_2^\mu\to a_2^\mu-\theta_2k^\mu\,.
\end{equation}
Then, the terms (\ref{w1and2}) have particularly simple gauge transformation properties and under (\ref{gtpol}) we have
\begin{equation}\label{wgt}
    \w{1}{\pbar}\to \w{1}{\pbar}-\theta_1\qquad \mathrm{while} \qquad \w{2}{\pbar}\to \w{2}{\pbar}-\theta_2\,.
\end{equation}
Hence, from (\ref{dist}) and (\ref{wgt}), we see that the Volkov field $\phiV(x)$ as defined in  (\ref{phivol}) will transform under a background field gauge transformation as expected for such a matter term:
\begin{equation}\label{phivgt}
    \phiV(x)\to\ee^{-ie\theta(x)}\phiV(x)\,.
\end{equation}

As in the free theory, we would like to interpret the Volkov field's modes that enter into (\ref{phivol}) as creation and annihilation operators and hence develop a description of the Volkov states and thence the Volkov propagator. In order to do this we need to be able to invert (\ref{phivol}) so that we can express the operators $\aV(p)$, etc, in terms of the Volkov fields. In the free case, this inversion is achieved using basic Fourier transform ideas, but now the distorted plane waves complicate things. However, simple physical guidelines will allow us to generalise the Fourier inversion to this Volkov description.

The guiding principles in finding an inversion for (\ref{phivol}) are that the Volkov modes are gauge invariant and time-independent. Given that the inversion result should reduce to the free field expression when the coupling is switched off, after a little experimentation we are lead to the expression:
\begin{equation}\label{av}
    \aV(p)=-i\int d^3x\,\big(\left(D_0\Dc(x,\pbar)\right)^\dag\phiV(x)-\Dc(x,\pbar)^\dag D_0\phiV(x)\big)\,,
\end{equation}
with similar expressions for the other modes.
In this equation gauge invariance is ensured through the use of the covariant derivative $D_0=\partial_0+ieA_0$ and time independence follows from the equations of motion (\ref{KGA}) and (\ref{disteqn}) along with the simple identity that $\partial_0D_0=D^2-D_iD^i-ieA_0D_0$. The actual proof of (\ref{av}), and its generalisation to the other modes and the associated creation operators  follows from the identity that
\begin{align}\label{avint}
    -i\int d^3x\,\big(\left(D_0\Dc(x,\pbar)\right)^\dag\phiV(x)&-\Dc(x,\pbar)^\dag D_0\phiV(x)\big)\\
    &=\intq\left(I(\pbar,\qbar)\aV(q)+I(\pbar,-\qbar)\bdV(q)\right)\,,\nonumber
\end{align}
where we have defined the integral
\begin{equation}\label{integral}
 I(\pbar,\qbar):=-i\int d^3x\,\left(\left(D_0\Dc(x,\pbar)\right)^\dag\Dc(x,\qbar)-\Dc(x,\pbar)^\dag D_0\Dc(x,\qbar)\right)\,.
\end{equation}
In  Appendix~\ref{appen2} we show that this can be  evaluated explicitly with the result that
\begin{equation}\label{intres}
    I(\pbar,\qbar)=(2\pi)^3(E^*_{p}+E^*_{q})\delta^3(\underline{p}-\underline{q})\,.
\end{equation}
The proof of the inversion result (\ref{av}) then follows immediately.

Given the expression (\ref{av}) for the Volkov modes and its adjoint
\begin{equation}\label{adv}
    \adV(q)=i\int d^3y\,\big(\left(D_0\Dc(y,\qbar)\right)\phiV^\dag(y)-\Dc(y,\qbar)\big( D_0\phiV(y)\big)^\dag\big)\,,
\end{equation}
we can exploit the time-independence of these results to calculate their commutator by using the equal-time canonical commutators of the Volkov fields. From this we see that
\begin{equation}\label{avcomm}
    [\aV(p),\adV(q)]=(2\pi)^32E^*_{p}\,\delta^3(\underline{p}-\underline{q})\,.
\end{equation}
Given these commutators we can now construct the unequal-time commutators:
\begin{equation}\label{uetcomm}
    [\phiV(x),\phiV^\dag(y)]=\intp\big(\Dc(x,\pbar)\Dc(y,\pbar)^\dag-\Dc(x,-\pbar)\Dc(y,-\pbar)^\dag\big)\,,
\end{equation}
the study of which lies at the heart of the causal properties of these fields and will be the subject of another publication. For now we shall consider   the Volkov states and their propagation.

From (\ref{avcomm})  we see that the Volkov modes satisfy the usual commutation relations which allows us to identify them as creation/annihilation operators, but with a relativistic normalisation given by the effective energy $E^*_{p}$. So acting on the Volkov vacuum, where $\aV(p)\ketV{0}=\bV(p)\ketV{0}=0$, we can construct multi-particle Volkov states. However, in contrast to the free case where $\phi^\dag(x)\ket{0}$ can be understood as a superposition of single particle states since $(\partial^2+m^2)\phi^\dag(x)\ket{0}=0$, for the Volkov field $\phiV^\dag(x)\ketV{0}$ is a much richer ensemble of states. As discussed in Appendix~A, the distortion term can be written as a superposition of Bessel functions and thus we can write
\begin{equation}\label{volsup}
    \phiV^\dag(x)\ketV{0}=\sum_{n=-\infty}^\infty \phi_n^\dag(x)\ketV{0}
\end{equation}
where
\begin{equation}\label{sidebandstates}
    \phi_n^\dag(x)\ketV{0}=\intp \ee^{i(\pbar-nk){\cdot}x}\ee^{-in\psi_\pbar}\mathrm{J}_n(ew_\pbar)\aV^\dag(p)\ketV{0}\,.
\end{equation}
These now satisfy a shifted free equation of motion:
\begin{equation}\label{sbandcond}
    ((\partial+ink)^2+m_*^2)\phi_n^\dag(x)\ketV{0}=0\,.
\end{equation}
We shall refer to them as sideband states in analogy with  Reiss'~\cite{Reiss:2009} discussion of the multi-pole structure of the Volkov propagator.

As is the case for the free theory, we expect the causal propagator for the Volkov field to be related to the two-point function
\begin{equation}\label{2poinv}
  \braV{0}\mathrm{T}\phiV(x)\phiV^\dagger(y)\ketV{0}\,,
\end{equation}
where the time ordering is given by
\begin{equation}\label{timeorder}
  \mathrm{T}\phiV(x)\phiV^\dagger(y)=\theta(x^0-y^0)\phiV(x)\phiV^\dagger(y)+\theta(y^0-x^0)\phiV^\dagger(y)\phiV(x)
\end{equation}
and $\theta(z)$ is the step function which we can write as
\begin{equation}\label{stepfn}
  \theta(z)=-\lim_{\epsilon\to 0^{+}}\int_{-\infty}^\infty{\bar{}\kern-0.45em d}\lambda\frac{i}{\lambda-i\epsilon}\ee^{i\lambda z}\,.
\end{equation}

\noindent The mode decomposition (\ref{volsup}) of the Volkov field will induce a double decomposition for the two-point function:
\begin{equation}\label{twopointbi}
  \braV{0}\mathrm{T}\phiV(x)\phiV^\dagger(y)\ketV{0}=\sum_{m,n}\braV{0}\mathrm{T}\phiV(x)\phiV^\dagger(y){\ketV{0}}^{\!\!\!\!\!\!mn}
: =\sum_{m,n} \braV{0}\mathrm{T}\phi_m(x)\phi_n^\dagger(y)\ketV{0}
\end{equation}
and this poses a challenge to the direct identification of the propagator.

We recall that in the free theory the field acting on the vacuum creates particle states which can then be annihilated at a later time by the conjugate field. This insight allows us to identify the propagator  as (the vacuum expectation value of) the time-ordered product of these fields. For the Volkov field we have seen that this particle interpretation is very different.  The modes of the field, $\phi^\dag_n(x)$, can be interpreted as creating particle states~(\ref{sidebandstates}). However,  each of these modes has its own on-shell type of condition~(\ref{sbandcond}). Thus the annihilation of such particles at a later time requires the corresponding conjugate mode. Therefore, in the sum~(\ref{twopointbi}), only the diagonal terms have an interpretation as a propagator. This identification is the operator version of the degeneracy argument given in the previous section. The interpretation  of the  off-diagonal terms in the two-point function is less clear but is related to the diagrams in Fig.~4 and will be addressed elsewhere.

From the commutation relations (\ref{avcomm}) and the expression (\ref{sidebandstates}) we thus identify the propagator as a sum over terms of the form
\begin{align}\label{propn1}
 \braV{0}\mathrm{T}\phi_n(x)\phi_n^\dagger(y)\ketV{0} =
  -\intp\,\,\, {\bar{}\kern-0.45em d}\lambda\frac{i}{\lambda-i\epsilon}&{\Big(}\ee^{i(\lambda-E^*_p+nk_0)(x^0-y^0)}\ \ee^{i(\underline{p}-n\underline{k}){\cdot}(\underline{x}-\underline{y})} \\
  \nonumber & +\ee^{-i(\lambda-E^*_p-nk_0)(x^0-y^0)}\ \ee^{-i(\underline{p}+n\underline{k}){\cdot}(\underline{x}-\underline{y})}{\Big)}\mathrm{J}^2_n(ew_\pbar)\,.
\end{align}
In order to make this look more like a propagator we need to go from on-shell expressions for the momenta to off-shell ones. The way this is done is intuitively clear but the details are a little subtle. Basically we want to identify the integration variable $\lambda$ as the missing four-momentum component $p^0$. To see how this works in practice, we note that this integral  is a sum of two integrals and we need to make different changes of variables in each. In the first integral we set $p^0=\lambda-E^*_p+nk^0$ while in the second $-p^0=\lambda-E^*_p-nk^0$ to give:
\begin{align}\label{propn2}
 \braV{0}\mathrm{T}\phi_n(x)\phi_n^\dagger(y)\ketV{0} =
  -\int
  \,\,\, {\bar{}\kern-0.45em d}^{\,4}p\,\ee^{ip^0(x^0-y^0)}{\Big(}&\frac{i}{p^0+E^*_p-nk^0-i\epsilon}\ \ee^{i(\underline{p}-n\underline{k}){\cdot}(\underline{x}-\underline{y})} \\
  \nonumber & +\frac{i}{E^*_p-p^0+nk^0-i\epsilon}\ \ee^{-i(\underline{p}+n\underline{k}){\cdot}(\underline{x}-\underline{y})}{\Big)}\frac1{2E^*_p}\mathrm{J}^2_n(ew_\pbar)\,.
\end{align}
We now shift the 3-momenta, and again we have a different shift in each term: in the first integral we change $\underline{p}\to -\underline{p}+n\underline{k}$ while in the second  $\underline{p}\to \underline{p}-n\underline{k}$.

In both cases (since $E_p^*=E_{-p}^*$) we get the shift $E_p^*\to E_{p-nk}^*=\sqrt{|\underline{p}-n\underline{k}|^2+m_*^{2}}$, the change to $w_\pbar$ are, however, a little more subtle as we now explain. Under our changes of variable the on-shell momenta change in different ways. In particular, under the first shift we get $\pbar\to (E^*_{-p+nk},  -\underline{p}+n\underline{k})$ while under the second we have $\pbar\to (E^*_{p-nk},  \underline{p}-n\underline{k})$. Both of these are on-shell but at different momenta. However, exploiting the fact that $w_\pbar=w_{-\pbar}$, we can view these shifts within $w_\pbar$ as
\begin{equation}\label{momentashift}
 \pbar\to \widetilde{p-nk}=(-E^*_{p-nk}, \underline{p}-n\underline{k})\qquad \mathrm{and} \qquad \pbar\to \overline{p-nk}=(E^*_{p-nk},\underline{p}-n\underline{k})\,.
\end{equation}
These are the negative and positive on-shell momenta corresponding to the on-shell condition $(p-nk)^2=m_*^{2}$.

Under these changes of variables, (\ref{propn2}) becomes (after putting terms over a common denominator)
\begin{align}\label{propn4}
 \braV{0}\mathrm{T}\phi_n(x)\phi_n^\dagger(y)\ketV{0} =&
  \int
  \,\,\, {\bar{}\kern-0.45em d}^{\,4}p\,\frac{i}{(p-nk)^2-m_*^{2}+i\epsilon}\ee^{ip{\cdot}(x-y)}\\&\times {\Big(}\frac{E^*_{p-nk}-(p^0-nk^0)}{2E^*_{p-nk}}\mathrm{J}^2_{n}(ew_{\widetilde{p-nk}})
  \nonumber  +\frac{E^*_{p-nk}+(p^0-nk^0)}{2E^*_{p-nk}}\mathrm{J}^2_{n}(ew_{\overline{p-nk}}){\Big)}\,.
\end{align}
We now note that the terms in brackets greatly simplify when evaluated at the two poles of the overall denominator where they become either $\mathrm{J}^2_n(ew_{\widetilde{p-nk}})$ or $\mathrm{J}^2_n(ew_{\overline{p-nk}})$. The off-shell version of these are both then $\mathrm{J}_n^2(ew_{p-nk})=\mathrm{J}_n^2(ew_p)$ since $a{\cdot}k=k^2=0$.  This allows us to write the full Volkov propagator in a fully off-shell form as (noting that $J_n^2(\alpha)=J_{-n}^2(\alpha)$)
\begin{equation}\label{fulloffshellV}
  \sum_{n=-\infty}^\infty \int
  \,\,\, {\bar{}\kern-0.45em d}^{\,4}p\,\frac{i\mathrm{J}^2_n(ew_p)}{(p+nk)^2-m_*^{2}+i\epsilon}\ee^{ip{\cdot}(x-y)}\,.
\end{equation}
From this expression we can deduce that the all-orders expression for the laser induced wave function renormalisation is
\begin{equation}\label{allorderz}
  Z^{(n)}_2(w_p)=\mathrm{J}^2_n(ew_p)\,.
\end{equation}
Using this identification and the standard expansions for the Bessel function, we see that to order $e^6$
\begin{align}\label{z2order6}
  \zn0(w_p) &=  1-e^2\frac{w_p^2}2+ e^4\frac{3w_p^4}{32}-e^6\frac{5w_p^6}{576}\,,\nonumber \\
  \zn{\pm1}(w_p) &= e^2\frac{ w_p^2}4-e^4\frac{w_p^4}{16}+e^6\frac{5w_p^6}{768}\,, \\
  \zn{\pm2}(w_p) &= e^4\frac{w_p^4}{64}-e^6\frac{w_p^6}{384}\,, \nonumber\\
  \zn{\pm3} (w_p)&= e^6\frac{w_p^6}{2304}\,.\nonumber
\end{align}
Here we see confirmation and an extension of the perturbative calculations presented in section 2. Indeed, the identification (\ref{allorderz}) allows us to conclude that to all orders,
\begin{equation}\label{zcparity}
  \zn{n}(w_p)=\zn{-n}(w_p)
\end{equation}

\section{Polarisation dependence}\label{four}
In this section we will consider the scalar matter in a linearly polarised laser background to explore the polarisation dependence of our results developed in the previous two sections. For a linearly polarised laser the background potential (\ref{cirpot}) becomes
\begin{equation}\label{lpot}
  A_\mu(x)=a_\mu\cos(\kdotx)\,,
\end{equation}
where the amplitude vector $a_\mu$ satisfies $k{\cdot} a=0$.
Following the techniques of Sect.~II, we find the Feynman rules shown in Fig.'s~\ref{lin4Frules261012}  and \ref{lin3Frules261012}.

\begin{figure}
\begin{minipage}[c]{1\linewidth}
\begin{tikzpicture}[xscale=0.6, yscale=0.75]
 \draw[electron] (0,0) -- node[below=4.5pt]{$p$}(2,0);
 \draw[laser] (0,1.5) -- (2,0);
 \draw[laser] (4,1.5) -- (2,0);
 \draw[electron] (2,0) -- node[below=2pt]{$p$}(4,0);
 \fill (2, 0) circle (1.5pt);
 \draw[white] (4,0) -- node [right=2pt]
{$\displaystyle\color{black}=
ie^2\frac{a^2}2\qquad$}(4,1.5);
\end{tikzpicture}
\begin{tikzpicture}[xscale=0.6, yscale=0.75]
 \draw[electron] (0,0) -- node[below=4.5pt]{$p$}(2,0);
 \draw[laser] (0,1.5) -- (2,0);
 \draw[laser] (0,0.8) -- (2,0);
 \draw[electron] (2,0) -- node[below=2pt]{$p+2k$}(4,0);
 \fill (2, 0) circle (1.5pt);
 \draw[white] (4,0) -- node [right=2pt]
{$\displaystyle\color{black}=$}(4,1.4);
\end{tikzpicture}
\begin{tikzpicture}[xscale=0.6, yscale=0.75]
 \draw[electron] (0,0) -- node[below=4.5pt]{$p$}(2,0);
 \draw[laser] (4,1.5) -- (2,0);
 \draw[laser] (4,0.8) -- (2,0);
 \draw[electron] (2,0) -- node[below=2pt]{$p-2k$}(4,0);
 \fill (2, 0) circle (1.5pt);
 \draw[white] (4,0) -- node [right=2pt]
{$\displaystyle\color{black}=
ie^2\frac{a^2}4$}(4,1.5);
\end{tikzpicture}
\end{minipage}
\caption{\label{lin4Frules261012}The four point Feynman rules for linear polarisation}
\end{figure}

\begin{figure}
\begin{minipage}[c]{0.6\linewidth}
\begin{tikzpicture}[xscale=0.6, yscale=0.75]
 \draw[electron] (0,0) -- node[below=4.5pt]{$p$}(2,0);
 \draw[laser] (0,1.5) -- (2,0);
 \draw[electron] (2,0) -- node[below=2pt]{$p+k$}(4,0);
 \fill (2, 0) circle (1.5pt);
 \draw[white] (4,0) -- node [right=2pt]
{$\displaystyle\color{black}=
$}(4,1.5);
\end{tikzpicture}
\begin{tikzpicture}[xscale=0.6, yscale=0.75]
 \draw[electron] (0,0) -- node[below=4.5pt]{$p$}(2,0);
 \draw[laser] (4,1.5) -- (2,0);
 \draw[electron] (2,0) -- node[below=2pt]{$p-k$}(4,0);
 \fill (2, 0) circle (1.5pt);
  \draw[white] (4,0) -- node [right=2pt]
{$\displaystyle\color{black}=
ie a\cdot p$}(4,1.5);
\end{tikzpicture}
\end{minipage}
\caption{\label{lin3Frules261012}The three point Feynman rules for linear polarisation}
\end{figure}

Comparing with Fig.'s~\ref{seagullcircrule} and \ref{threecircrule} we see that the Feynman rules are polarisation dependent. The most important difference is the existence of additional four point vertices where both photons are incoming or both outgoing. At order $e^2$ these rules do not generate diagrams in the propagator as the initial and final matter momenta would not be the same. Thus the propagator at this order is still given by the degenerate diagrams of Fig.'s~\ref{seagullcircrule} and \ref{line2propdiags}. Thus the mass shift is now given by
\beq\label{eqn131212a}
\delta m^2=-\frac{e^2 a^2}2\,,
\eeq
where the factor of a half follows from the difference between the Feynman rules in this  case. Here we see the polarisation dependence of the mass shift which is an example of the dependence of the mass on the form of the laser field~\cite{Harvey:2012ie}.

\begin{figure}
\begin{minipage}[c]{0.25\linewidth}
\begin{tikzpicture}[xscale=0.6, yscale=0.75]
 \draw[electron] (0,0) -- node[below=4.5pt]{$p$}(2,0);
 \draw[electron] (2,0) -- node[below=4.5pt]{$p+2k$}(4,0);
\draw[electron] (4,0) -- node[below=4.5pt]{$p$}(6,0);
 \draw[laser] (0,1.5) -- (2,0);
\draw[laser] (0,0.8) -- (2,0);
\fill[white] (3.02,0.78) circle (2.5pt);
 \fill (2, 0) circle (1.5pt);
\draw[laser] (6,1.5) -- (4,0);
 \draw[laser] (6,0.8) -- (4,0);
 \fill (4, 0) circle (1.5pt);
\end{tikzpicture}
\end{minipage}
\quad
\begin{minipage}[c]{0.25\linewidth}
\begin{tikzpicture}[xscale=0.6, yscale=0.75]
\draw[electron] (0,0) -- node[below=4.5pt]{$p$}(2,0);
 \draw[electron] (2,0) -- node[below=4.5pt]{$p+2k$}(4,0);
 \draw[electron] (4,0) -- node[below=4.5pt]{$p+k$}(6,0);
 \draw[electron] (6,0) -- node[below=4.5pt]{$p$}(8,0);
 \draw[laser] (0,1.5) -- (2,0);
\draw[laser] (0,0.8) -- (2,0);
 \fill[white] (4,0.78) circle (2.5pt);
 \fill (2, 0) circle (1.5pt);
 \fill[white] (5.02,0.78) circle (2.5pt);
 \draw[laser] (6.1,1.5) -- (4,0);
 \draw[laser](8.1,1.5) -- (6,0);
 \fill (2, 0) circle (1.5pt);
 \fill (4,0) circle (1.5pt);
 \fill (6, 0) circle (1.5pt);
\end{tikzpicture}
\end{minipage}
\caption{\label{swingingseagulls}Examples of additional, higher order, degenerate diagrams in a linearly polarised laser}
\end{figure}
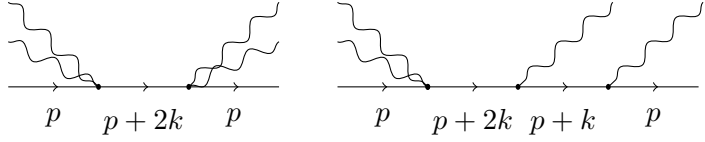

The wave function renormalisations similarly are
\beq
\zn0 =1-e^2\frac{u_p}2\,.
\eeq
where
\beq
u_p=-\frac{p\cdot a}{p\cdot k}\,.
\eeq
and
\beq
\zn{\pm1} = e^2\frac{u^2}{4}
\,.
\eeq
At order $e^4$ there are twenty one degenerate diagrams that contribute to the propagator. These are those of Fig.~\ref{nnl} plus an additional eight diagrams
which involve the new seagull Feynman rules. These include, for example, the diagrams of Fig.~\ref{swingingseagulls}.

We will simply summarise the results obtained for the propagator up to order $e^4$. As with the circular polarisation case, we find that the mass shift is not altered by higher order diagrams. The result is still that of~(\ref{eqn131212a}). However, for the wave function renormalisations, with notation being a simple extension of the circular polarisation case, the results are rather different. Defining
\begin{equation}\label{vdef}
  v_p=\frac{a^2}{8p\cdot k}\,,
\end{equation}
we obtain
\beq\label{z0l}
\zn0(u_p,v_p) =1-e^2\frac{u^2_p}2- e^4\left(\frac{v_p^2}{2}-  \frac{3u_p^4}{32}\right)
\,,
\eeq
and
\beq
\zn{\pm1}(u_p,v_p) = e^2\frac{u_p^2}{4}- e^4\left( \frac{u_p^4}{16}\pm \frac{v_pu_p^2}{4}\right)
\,,
\eeq
while
\beq\label{z2l}
\zn{\pm2}(u_p,v_p) =e^4\left(\frac{u_p^4}{64}+  \frac{v_p^2}{4} \pm  \frac{v_pu_p^2}{8}\right)
\,.
\eeq
Again we note that $\zn{\pm n}$ is of order $e^{2n}$ and higher. Furthermore we see that although once again only the seagull vertex produces contributions to the mass shift, the wave function renormalisation for linear polarisations includes contributions from both seagull and three point vertices. Thus  the fact that the wave function renormalisation for circular polarised lasers was generated by three point vertices alone is demonstrated to be a polarisation artefact. We also note that in contrast to the circular polarisation case, $\zn{n}$ does not equal $\zn{-n}$.

At the operator level, the distortion term for a linearly polarised background is
\begin{align}
  \Dl(x,\pbar)&=\ee^{-i\pbar{\cdot}x}\ee^{i(eu_\pbar\sin(\kdotx)+e^2v_\pbar\sin(2\kdotx))} \nonumber \\
  & =\ee^{-i\pbar{\cdot}x}\sum_{n=-\infty}^\infty \ee^{in\kdotx}{\mathrm{J}}_n(eu_p,e^2v_p)\,.
\end{align}
An important point to note about this expression, and one that did not arise for circular polarisation, is that the arguments of the generalised Bessel function are not invariant under the replacement $\pbar \to -\pbar$. In particular, although $u_{-\pbar}=u_\pbar$, we now have from (\ref{vdef}) that $v_{-\pbar}=-v_\pbar$. This means that in the derivation of the propagator for a linearly polarised background, we need to replace equation (\ref{propn1}) by
\begin{align}\label{propnl1}
 \braV{0}\mathrm{T}\phi_n(x)\phi_n^\dagger(y)\ketV{0} =
  -\intp\,\,\, {\bar{}\kern-0.45em d}\lambda&\frac{i}{\lambda-i\epsilon}{\Big(}\ee^{i(\lambda-E^*_p+nk_0)(x^0-y^0)}\ \ee^{i(\underline{p}-n\underline{k}){\cdot}(\underline{x}-\underline{y})} \Jb_n^2(eu_\pbar,e^2v_\pbar) \\
  \nonumber & +\ee^{-i(\lambda-E^*_p-nk_0)(x^0-y^0)}\ \ee^{-i(\underline{p}+n\underline{k}){\cdot}(\underline{x}-\underline{y})}\Jb_n^2(eu_\pbar,-e^2v_\pbar){\Big)}\,.
\end{align}

If we now apply the same arguments as before for circular polarisation, we obtain in place of (\ref{propn4}) the expression
\begin{align}\label{propnl4}
 \braV{0}&\mathrm{T}\phi_n(x)\phi_n^\dagger(y)\ketV{0} =
  \int
  \,\,\, {\bar{}\kern-0.45em d}^{\,4}p\,\frac{i}{(p-nk)^2-m_*^{2}+i\epsilon}\ee^{ip{\cdot}(x-y)}\\&\times {\Big(}\frac{E^*_{p-nk}-(p^0-nk^0)}{2E^*_{p-nk}}\mathrm{J}^2_{n}(eu_{\widetilde{p-nk}},-e^2v_{\widetilde{p-nk}})
  \nonumber  +\frac{E^*_{p-nk}+(p^0-nk^0)}{2E^*_{p-nk}}\mathrm{J}^2_{n}(eu_{\overline{p-nk}},-e^2v_{\overline{p-nk}}){\Big)}\,.
\end{align}
Repeating the argument under (\ref{propn4}) and using the identity (\ref{bessid}), we obtain the propagator in a  linearly polarised background
\begin{equation}\label{fulloffshelllV}
  \sum_{n=-\infty}^\infty \int
  \,\,\, {\bar{}\kern-0.45em d}^{\,4}p\,\frac{i\mathrm{J}^2_n(eu_p,e^2v_p)}{(p+nk)^2-m_*^{2}+i\epsilon}\ee^{ip{\cdot}(x-y)}\,.
\end{equation}
Hence the wave function renormalisation constants in this linearly polarised background for all the states at all orders are given by
\begin{equation}\label{wavefunlin}
  \zn{n}(eu_p,e^2v_p)=\mathrm{J}^2_n(eu_p,e^2v_p)\,.
\end{equation}
It is easy to verify that this agrees with the perturbative results (\ref{z0l}) to (\ref{z2l}).


\section{Conclusions}


In this paper we have calculated the propagator of an electron in a laser background in two different ways as well
as in different laser backgrounds. For simplicity this work was carried out in scalar QED. The first approach we pursued was motivated by infra-red physics: the physical propagator is composed of the sum of all degenerate (indistinguishable) processes. For example, an electron which passes freely through a laser is indistinguishable from one which, say, absorbs a laser photon and emits a photon of the same energy and momentum back into the laser. Thus the propagator in such a background should involve a sum of tree diagrams with Fig.'s~\ref{seagullcircrule} and \ref{line2propdiags} showing the order $e^2$ degenerate processes.

We have thus shown that this leads to the well known mass shift and also that the perturbative corrections generate additional poles (sideband states) in the propagator of the form
\begin{equation}\label{15121356}
  \frac1{(p\pm nk)^2-m_*^{2}}\,,
\end{equation}
for integer $n$. We stress that sideband states arise perturbatively~\cite{Reiss:2009}, some already from the lowest order corrections to the free propagator. The wave function renormalisation factors, i.e., the numerators of these sideband poles, start at order $e^{2n}$ (with $n$ as above). In this way we have calculated the degenerate corrections to the free propagator up to next to leading order.

In the scalar theory we consider here, there are both three and four point interactions between the matter and laser fields. The form of these Feynman rules depends upon the laser polarisation. However, for both circularly and linearly polarised lasers the mass shift was solely generated by contributions from the four-point vertex at lowest order. All higher order diagrams which generate double poles simply yield the higher order terms in the geometric expansion~(\ref{eqnHRY2343}). It was also shown that the mass shift in all poles was the same.

The wave function renormalisations for circularly polarised lasers only involve the three point vertex structures. There is also the symmetry that $\zn{n}=\zn{-n}$. However, for linearly polarised lasers, the wave function renormalisation structures depended upon both sets of vertices. Another new feature of the linearly polarised case is that $\zn{n}\neq\zn{-n}$, i.e., the symmetry between these sideband states visible for circularly polarised laser backgrounds is removed for linear polarisation.

The second approach which we have pursued in this paper was based on the canonical quantisation of these fields. We are not aware of such an investigation in the literature although we note~\cite{Boca:2011m} and some  lightcone quantisation work~\cite{Neville:1971uc,Ilderton:2012qe}. Our approach yielded an all orders result which completely agreed with the Feynman diagram results. We have constructed the canonical quantisation for Volkov states including their unequal time commutators. The two-point function for Volkov fields was written as a double sum over creation and annihilation type modes. These modes satisfy a free equation of motion, see~(\ref{sbandcond}), but shifted in accord with the sideband states discussed above. However, as only the diagonal terms in the sum~(\ref{twopointbi}) have a propagator interpretation we restricted our attention to these terms. The Volkov propagator defined in this way could be expressed as a sum over sideband states and the normalisation given in terms of Bessel functions (generalised Bessel functions for linearly polarised laser backgrounds) see~(\ref{allorderz}) and~(\ref{wavefunlin}). Expanding these results yielded all of our previous perturbative calculations. It is not immediately obvious to us that our infra-red motivated sum over degenerate diagrams and the diagonal terms in the canonical quantisation double sum had to agree like this.

It is useful to recall here that, even without a background field,  the pole in the free propagator is modified in perturbative calculations to yield the full spectral decomposition of the two-point function. This now includes renormalised poles, bound states and branch cuts corresponding to scattering states. What we have shown in this paper, without calculating any loop corrections,  is that interactions with the background induce a modified spectral structure with infinitely many discrete poles corresponding to the sideband states. It is intriguing to speculate how the loop corrections will further modify these sideband structures. In particular, one might ask whether infra-red divergences will prevent sideband states from displaying mass gaps and, following Lee-Nauenberg, what real processes should be summed over to construct finite inclusive cross-sections in a laser background.

This work raises many further questions such as the extension to fermionic QED, the role of the off-diagonal two-point function structures and the general form of the vertices both perturbatively and non-perturbatively. They will be addressed in future publications.

\appendix

\section{Generalisations of Bessel functions}
The distortion of the plane wave solutions due to the interaction with the laser are best described using Bessel functions. For a circularly polarised laser there are two ways to formulate this. One is in terms of standard Bessel functions while in the other a generalised formulation of Bessel functions is introduced. It is useful to use  both of these and understand how to move between them. For a linearly polarised laser the only option is  the generalised route and this will be easier to understand once we have discussed the circular case.
\subsection{Circular Polarisation}
In the expression (\ref{dist}) for the distortion term $\Dc(x,\pbar)$ we can write
\begin{equation}\label{bes1}
    \w{1}{\pbar}\sin(\kdotx)+\w{2}{\pbar}\cos(\kdotx)=\w{}{\pbar}\sin(\kdotx+\psi_{\pbar})
\end{equation}
where
\begin{equation}\label{bes2}
    \w{1}{\pbar}^2+\w{2}{\pbar}^2=\w{}{\pbar}^2\qquad \mathrm{and}\qquad \tan\psi_{\pbar}=\frac{\w{2}{\pbar}}{\w{1}{\pbar}}\,.
\end{equation}
Using this we can then write the distortion term in terms of standard Bessel functions:
\begin{equation}\label{bes3}
    \Dc(x,\pbar)=\ee^{-i\pbar{\cdot}x}\sum_{n=-\infty}^\infty \ee^{in(\kdotx+\psi_{\pbar})}\mathrm{J}_n(e\w{}{\pbar})\,.
\end{equation}
The advantage of this formulation is that in the propagator we will always have even powers of $\w{}{\pbar}$ and the phase $\psi_{\pbar}$ will cancel so that any ambiguity in the sign of $\w{}{\pbar}$ can be avoided. But, in deriving the inversion formula (\ref{intres}), odd powers of $\w{}{\pbar}$ will arise and it is much simpler to use a generalisation of the usual Bessel functions as outlined below.

Rather than rewriting the sine and cosine terms in $\Dc(x,\pbar)$ as a shifted sine as in (\ref{bes1}) and then using the Bessel representation, we can directly express both the sine and cosine terms as Bessel functions to arrive at the representation
\begin{equation}\label{bes4}
    \Dc(x,\pbar)=\ee^{-i\pbar{\cdot}x}\sum_{n=-\infty}^\infty \ee^{in\kdotx}\Jt_n(e\w{1}{\pbar},e\w{2}{\pbar})\,,
\end{equation}
where we have defined the generalised Bessel function $\Jt_n(\alpha,\beta)$ by
\begin{equation}\label{bes5}
    \Jt_n(\alpha,\beta)=\sum_{r=-\infty}^\infty i^r \mathrm{J}_{n-r}(\alpha)\mathrm{J}_r(\beta)\,.
\end{equation}
These Bessel functions satisfy various identities that follow quite easily from their definition and we note, in particular, that $\Jt_n(0,0)=\delta_{n0}$ and, most importantly
\begin{equation}\label{bes6}
    2n\Jt_n(\alpha,\beta)=\alpha(\Jt_{n-1}(\alpha,\beta)+\Jt_{n+1}(\alpha,\beta))
    +i\beta(\Jt_{n-1}(\alpha,\beta)-\Jt_{n+1}(\alpha,\beta))\,.
\end{equation}
As we shall see, both (\ref{bes3}) and (\ref{bes4}) are useful representations of the distortion factor and we can shift between them by using the identity
\begin{equation}\label{bes7}
    \ee^{in\psi_{\pbar}}\mathrm{J}_n(e\w{}{\pbar})=\Jt_n(e\w{1}{\pbar},e\w{2}{\pbar})\,.
\end{equation}
\subsection{Linear Polarisation}
In the linear case the only option is to expand both the $\sin(\kdotx)$ and the $\sin(2\kdotx)$ terms in $\Dl(x,\pbar)$ as series of Bessel functions to give the generalised Bessel function expansion (see Appendix J of \cite{Krainov:book})
\begin{equation}\label{bes8}
    \Dl(x,\pbar)=\ee^{-i\pbar{\cdot}x}\sum_{n=-\infty}^\infty \ee^{in\kdotx}\Jb_n(eu_{\pbar},e^2v_{\pbar})\,,
\end{equation}
where now the generalised Bessel functions $\Jb_n(\alpha,\beta)$ are defined by
\begin{equation}\label{bes9}
    \Jb_n(\alpha,\beta)=\sum_{r=-\infty}^\infty \mathrm{J}_{n-2r}(\alpha)\mathrm{J}_r(\beta)\,.
\end{equation}
Again, these Bessel functions also satisfy $\Jb_n(0,0)=\delta_{n0}$ and now
\begin{equation}\label{bes10}
    2n\Jb_n(\alpha,\beta)=\alpha(\Jb_{n-1}(\alpha,\beta)+\Jb_{n+1}(\alpha,\beta))
    +2\beta(\Jb_{n-1}(\alpha,\beta)+\Jb_{n+1}(\alpha,\beta))\,.
\end{equation}
We also note that the generalised Bessel functions satisfy the identity
\begin{equation}\label{bessid}
  J_n(\alpha,-\beta)=(-1)^n J_{-n}(\alpha,\beta)\,.
\end{equation}

\section{Volkov Inversion Formula}
Here we shall derive (\ref{intres}) for circular polarised Volkov fields, the linear polarised result follows in much the same way (see (\cite{Merfat}) for the full details) and note that, for simplicity,  in this section we are setting the coupling to unity.

The first observation to make is that the gauge invariance of (\ref{integral}) can be exploited to work in the gauge where, in addition to $k{\cdot} A=0$, we have $A_0=A_3=0$. This means that the covariant derivatives in (\ref{integral}) can be replace by ordinary derivatives.
Using (\ref{dist}) it is then straightforward to see that
\begin{equation}\label{inv1}
    I(\pbar,\qbar)=\int d^3x\Big((E_{p}^*+E_{q}^*)-k_0(\w{1}{\pbar}+\w{1}{\qbar})\cos(\kdotx)+k_0(\w{2}{\pbar}+
    \w{2}{\qbar})\sin(\kdotx)\Big)\Dc(x,\pbar)^\dag \Dc(x,\qbar)\,.
\end{equation}
Writing the final product in terms of the generalised Bessel functions (\ref{bes5})
\begin{equation}\label{inv2}
    \Dc(x,\pbar)^\dag \Dc(x,\qbar)=\ee^{i(\pbar-\qbar){\cdot}x}\sum_{n=-\infty}^\infty \ee^{in\kdotx}
    \Jt_n(\w{1}{\qbar}-\w{1}{\pbar},\w{2}{\qbar}-\w{2}{\pbar})\,,
\end{equation}
and using the standard exponential description of the sine and cosine terms which allows us to perform the $x$-integral, we arrive at the expression
\begin{align*}
    I(\pbar,\qbar)=(2\pi)^3\sum_{n=-\infty}^\infty \delta^3(\underline{p}-\underline{q}+n\underline{k})
    &\ee^{i(E^*_{p}-E^*_{q}+nk_0)x^0}\\
    \times\Big((E^*_{p}+E^*_{q})\Jt_n-\tfrac12k_0(\w{1}{\pbar}+&\w{1}{\qbar})(\Jt_{n-1}+\Jt_{n+1})
    -\tfrac{i}2k_0(\w{2}{\pbar}+\w{2}{\qbar})(\Jt_{n-1}-\Jt_{n+1})\Big)\,,
\end{align*}
where the arguments of the Bessel functions have been suppressed but are all the same differences as  given in (\ref{inv2}). We now need to consider separately the cases $n=0$ and $n\ne0$.

When $n=0$ the delta function implies that $E^*_{p}=E^*_{q}$ and hence $\w{1}{\pbar}=\w{1}{\qbar}$ as well as  $\w{2}{\pbar}=\w{2}{\qbar}$, which means that the arguments of the Bessel functions vanish. Now using the result that $\Jt_n(0,0)=\delta_{n0}$ we get the contribution to $I(\pbar,\qbar)$ of the term  $(2\pi)^3 \delta^3(\underline{p}-\underline{q})(E^*_{p}+E^*_{q})$ which is the right hand side of the identity (\ref{intres}) that we wish to show. Hence we now need to verify that when $n\ne0$ we get no further contributions to $I(\pbar,\qbar)$.

When $n\ne0$ the delta function and gauge condition  imply that both $\w{1}{\pbar}\ne\w{1}{\qbar}$ and  $\w{2}{\pbar}\ne\w{2}{\qbar}$ so we can write the sum as
\begin{align}\label{inv3}
    \sum_{n\ne0} \delta^3(\underline{p}-\underline{q}+n\underline{k})
    \ee^{i(E^*_{p}-E^*_{q}+nk_0)x^0}\nonumber&\\
    \times\Big((E^*_{p}+E^*_{q})\Jt_n-\tfrac12k_0&
    \left(\frac{\w{1}{\pbar}+\w{1}{\qbar}}{\w{1}{\qbar}-\w{1}{\pbar}}\right)(\w{1}{\qbar}-\w{1}{\pbar})
    (\Jt_{n-1}+\Jt_{n+1})\\\nonumber
    -&\tfrac{1}2k_0\left(\frac{\w{2}{\pbar}+\w{2}{\qbar}}{\w{2}{\qbar}-\w{2}{\pbar}}\right)
    i(\w{2}{\qbar}-\w{2}{\pbar})(\Jt_{n-1}-\Jt_{n+1})\Big)\,.
\end{align}
The delta function in this also implies that both $\pbar{\cdot}a_1=\qbar{\cdot}a_1$ and $\pbar{\cdot}a_2=\qbar{\cdot}a_2$. Using this we see that
\begin{equation}\label{inv4}
    \frac{\w{1}{\pbar}+\w{1}{\qbar}}{\w{1}{\qbar}-\w{1}{\pbar}}=
    \frac{\w{2}{\pbar}+\w{2}{\qbar}}{\w{2}{\qbar}-\w{2}{\pbar}}=
    \frac{\qbar{\cdot}k+\pbar{\cdot}k}{\pbar{\cdot}k-\qbar{\cdot}k}=
    \frac{E^*_{p}+E^*_{q}-2p^3-nk_0}{E^*_{p}-E^*_{q}+nk_0}
    \,.
\end{equation}
Using this and the identity (\ref{bes6}) we can write the sum (\ref{inv3}) as
\begin{equation}\label{inv5}
    \sum_{n\ne0} \delta^3(\underline{p}-\underline{q}+n\underline{k})
    \ee^{i(E^*_{}-E^*_{q}+nk_0)x^0}
    \Big((E^*_{p}+E^*_{q})-\tfrac12k_0
    \left(\frac{E^*_{p}+E^*_{q}-2p^3-nk_0}{E^*_{p}-E^*_{q}+nk_0}\right) 2n\Big)\Jt_n\,.
\end{equation}
Putting the bracketed  terms over a common denominator quickly shows that this sum vanishes as required.
 \label{appen2}


%



\begin{acknowledgments}
 We wish to thank Chris Harvey, Tom Heinzl and Anton Ilderton for helpful discussions.
\end{acknowledgments}


\begin{thebibliography}{34}%
\makeatletter
\providecommand \@ifxundefined [1]{%
 \@ifx{#1\undefined}
}%
\providecommand \@ifnum [1]{%
 \ifnum #1\expandafter \@firstoftwo
 \else \expandafter \@secondoftwo
 \fi
}%
\providecommand \@ifx [1]{%
 \ifx #1\expandafter \@firstoftwo
 \else \expandafter \@secondoftwo
 \fi
}%
\providecommand \natexlab [1]{#1}%
\providecommand \enquote  [1]{``#1''}%
\providecommand \bibnamefont  [1]{#1}%
\providecommand \bibfnamefont [1]{#1}%
\providecommand \citenamefont [1]{#1}%
\providecommand \href@noop [0]{\@secondoftwo}%
\providecommand \href [0]{\begingroup \@sanitize@url \@href}%
\providecommand \@href[1]{\@@startlink{#1}\@@href}%
\providecommand \@@href[1]{\endgroup#1\@@endlink}%
\providecommand \@sanitize@url [0]{\catcode `\\12\catcode `\$12\catcode
  `\&12\catcode `\#12\catcode `\^12\catcode `\_12\catcode `\%12\relax}%
\providecommand \@@startlink[1]{}%
\providecommand \@@endlink[0]{}%
\providecommand \url  [0]{\begingroup\@sanitize@url \@url }%
\providecommand \@url [1]{\endgroup\@href {#1}{\urlprefix }}%
\providecommand \urlprefix  [0]{URL }%
\providecommand \Eprint [0]{\href }%
\providecommand \doibase [0]{http://dx.doi.org/}%
\providecommand \selectlanguage [0]{\@gobble}%
\providecommand \bibinfo  [0]{\@secondoftwo}%
\providecommand \bibfield  [0]{\@secondoftwo}%
\providecommand \translation [1]{[#1]}%
\providecommand \BibitemOpen [0]{}%
\providecommand \bibitemStop [0]{}%
\providecommand \bibitemNoStop [0]{.\EOS\space}%
\providecommand \EOS [0]{\spacefactor3000\relax}%
\providecommand \BibitemShut  [1]{\csname bibitem#1\endcsname}%
\let\auto@bib@innerbib\@empty
\bibitem [{\citenamefont {Volkov}(1935)}]{Volkov:1935zz}%
  \BibitemOpen
  \bibfield  {author} {\bibinfo {author} {\bibfnamefont {D.~M.}\ \bibnamefont
  {Volkov}},\ }\href {\doibase 10.1007/BF01331022} {\bibfield  {journal}
  {\bibinfo  {journal} {Z.Phys.}\ }\textbf {\bibinfo {volume} {94}},\ \bibinfo
  {pages} {250} (\bibinfo {year} {1935})}\BibitemShut {NoStop}%
\bibitem [{\citenamefont {Maiman}(1960)}]{Maiman:1960}%
  \BibitemOpen
  \bibfield  {author} {\bibinfo {author} {\bibfnamefont {T.~H.}\ \bibnamefont
  {Maiman}},\ }\href {\doibase 10.1038/187493a0} {\bibfield  {journal}
  {\bibinfo  {journal} {Nature}\ }\textbf {\bibinfo {volume} {187}},\ \bibinfo
  {pages} {493} (\bibinfo {year} {1960})}\BibitemShut {NoStop}%
\bibitem [{\citenamefont {Reiss}(1962)}]{reiss:387}%
  \BibitemOpen
  \bibfield  {author} {\bibinfo {author} {\bibfnamefont {H.~R.}\ \bibnamefont
  {Reiss}},\ }\href {\doibase 10.1063/1.1724238} {\bibfield  {journal}
  {\bibinfo  {journal} {J. Math. Phys.}\ }\textbf {\bibinfo {volume} {3}},\
  \bibinfo {pages} {387} (\bibinfo {year} {1962})}\BibitemShut {NoStop}%
\bibitem [{\citenamefont {Nikishov}\ and\ \citenamefont
  {Ritus}(1964{\natexlab{a}})}]{Nikishov:1964zza}%
  \BibitemOpen
  \bibfield  {author} {\bibinfo {author} {\bibfnamefont {A.}~\bibnamefont
  {Nikishov}}\ and\ \bibinfo {author} {\bibfnamefont {V.}~\bibnamefont
  {Ritus}},\ }\href@noop {} {\bibfield  {journal} {\bibinfo  {journal}
  {Sov.Phys.JETP}\ }\textbf {\bibinfo {volume} {19}},\ \bibinfo {pages} {529}
  (\bibinfo {year} {1964}{\natexlab{a}})}\BibitemShut {NoStop}%
\bibitem [{\citenamefont {Nikishov}\ and\ \citenamefont
  {Ritus}(1964{\natexlab{b}})}]{Nikishov:1964zz}%
  \BibitemOpen
  \bibfield  {author} {\bibinfo {author} {\bibfnamefont {A.}~\bibnamefont
  {Nikishov}}\ and\ \bibinfo {author} {\bibfnamefont {V.}~\bibnamefont
  {Ritus}},\ }\href@noop {} {\bibfield  {journal} {\bibinfo  {journal}
  {Sov.Phys.JETP}\ }\textbf {\bibinfo {volume} {19}},\ \bibinfo {pages} {1191}
  (\bibinfo {year} {1964}{\natexlab{b}})}\BibitemShut {NoStop}%
\bibitem [{\citenamefont {Brown}\ and\ \citenamefont
  {Kibble}(1964)}]{Brown:1964zz}%
  \BibitemOpen
  \bibfield  {author} {\bibinfo {author} {\bibfnamefont {L.~S.}\ \bibnamefont
  {Brown}}\ and\ \bibinfo {author} {\bibfnamefont {T.~W.~B.}\ \bibnamefont
  {Kibble}},\ }\href {\doibase 10.1103/PhysRev.133.A705} {\bibfield  {journal}
  {\bibinfo  {journal} {Phys. Rev. A}\ }\textbf {\bibinfo {volume} {133}},\
  \bibinfo {pages} {705} (\bibinfo {year} {1964})}\BibitemShut {NoStop}%
\bibitem [{\citenamefont {Goldman}(1964)}]{Goldman1964103}%
  \BibitemOpen
  \bibfield  {author} {\bibinfo {author} {\bibfnamefont {I.}~\bibnamefont
  {Goldman}},\ }\href {\doibase 10.1016/0031-9163(64)90728-0} {\bibfield
  {journal} {\bibinfo  {journal} {Physics Letters}\ }\textbf {\bibinfo {volume}
  {8}},\ \bibinfo {pages} {103 } (\bibinfo {year} {1964})}\BibitemShut
  {NoStop}%
\bibitem [{\citenamefont {Bula}\ \emph {et~al.}(1996)\citenamefont {Bula} \emph
  {et~al.}}]{PhysRevLett.76.3116}%
  \BibitemOpen
  \bibfield  {author} {\bibinfo {author} {\bibfnamefont {C.}~\bibnamefont
  {Bula}} \emph {et~al.},\ }\href {\doibase 10.1103/PhysRevLett.76.3116}
  {\bibfield  {journal} {\bibinfo  {journal} {Phys. Rev. Lett.}\ }\textbf
  {\bibinfo {volume} {76}},\ \bibinfo {pages} {3116} (\bibinfo {year}
  {1996})}\BibitemShut {NoStop}%
\bibitem [{\citenamefont {Harvey}\ \emph {et~al.}(2009)\citenamefont {Harvey},
  \citenamefont {Heinzl},\ and\ \citenamefont {Ilderton}}]{Harvey:2009ry}%
  \BibitemOpen
  \bibfield  {author} {\bibinfo {author} {\bibfnamefont {C.}~\bibnamefont
  {Harvey}}, \bibinfo {author} {\bibfnamefont {T.}~\bibnamefont {Heinzl}}\
  and\ \bibinfo {author} {\bibfnamefont {A.}~\bibnamefont {Ilderton}},\ }\href
  {\doibase 10.1103/PhysRevA.79.063407} {\bibfield  {journal} {\bibinfo
  {journal} {Phys.Rev.}\ }\textbf {\bibinfo {volume} {A79}},\ \bibinfo {pages}
  {063407} (\bibinfo {year} {2009})},\ \Eprint {http://arxiv.org/abs/0903.4151}
  {arXiv:0903.4151 [hep-ph]} \BibitemShut {NoStop}%
\bibitem [{\citenamefont {Boca}\ and\ \citenamefont
  {Florescu}(2009)}]{Boca:2009zz}%
  \BibitemOpen
  \bibfield  {author} {\bibinfo {author} {\bibfnamefont {M.}~\bibnamefont
  {Boca}}\ and\ \bibinfo {author} {\bibfnamefont {V.}~\bibnamefont
  {Florescu}},\ }\href {\doibase 10.1103/PhysRevA.80.053403} {\bibfield
  {journal} {\bibinfo  {journal} {Phys.Rev.}\ }\textbf {\bibinfo {volume}
  {A80}},\ \bibinfo {pages} {053403} (\bibinfo {year} {2009})}\BibitemShut
  {NoStop}%
\bibitem [{\citenamefont {Heinzl}\ \emph {et~al.}(2010)\citenamefont {Heinzl},
  \citenamefont {Seipt},\ and\ \citenamefont {Kampfer}}]{Heinzl:2009nd}%
  \BibitemOpen
  \bibfield  {author} {\bibinfo {author} {\bibfnamefont {T.}~\bibnamefont
  {Heinzl}}, \bibinfo {author} {\bibfnamefont {D.}~\bibnamefont {Seipt}}\
  and\ \bibinfo {author} {\bibfnamefont {B.}~\bibnamefont {Kampfer}},\ }\href
  {\doibase 10.1103/PhysRevA.81.022125} {\bibfield  {journal} {\bibinfo
  {journal} {Phys.Rev.}\ }\textbf {\bibinfo {volume} {A81}},\ \bibinfo {pages}
  {022125} (\bibinfo {year} {2010})},\ \Eprint {http://arxiv.org/abs/0911.1622}
  {arXiv:0911.1622 [hep-ph]} \BibitemShut {NoStop}%
\bibitem [{\citenamefont {Mackenroth}\ and\ \citenamefont
  {Di~Piazza}(2011)}]{Mackenroth:2010jr}%
  \BibitemOpen
  \bibfield  {author} {\bibinfo {author} {\bibfnamefont {F.}~\bibnamefont
  {Mackenroth}}\ and\ \bibinfo {author} {\bibfnamefont {A.}~\bibnamefont
  {Di~Piazza}},\ }\href {\doibase 10.1103/PhysRevA.83.032106} {\bibfield
  {journal} {\bibinfo  {journal} {Phys.Rev.}\ }\textbf {\bibinfo {volume}
  {A83}},\ \bibinfo {pages} {032106} (\bibinfo {year} {2011})},\ \Eprint
  {http://arxiv.org/abs/1010.6251} {arXiv:1010.6251 [hep-ph]} \BibitemShut
  {NoStop}%
\bibitem [{\citenamefont {Seipt}\ and\ \citenamefont
  {Kampfer}(2011)}]{Seipt:2010ya}%
  \BibitemOpen
  \bibfield  {author} {\bibinfo {author} {\bibfnamefont {D.}~\bibnamefont
  {Seipt}}\ and\ \bibinfo {author} {\bibfnamefont {B.}~\bibnamefont
  {Kampfer}},\ }\href {\doibase 10.1103/PhysRevA.83.022101} {\bibfield
  {journal} {\bibinfo  {journal} {Phys.Rev.}\ }\textbf {\bibinfo {volume}
  {A83}},\ \bibinfo {pages} {022101} (\bibinfo {year} {2011})},\ \Eprint
  {http://arxiv.org/abs/1010.3301} {arXiv:1010.3301 [hep-ph]} \BibitemShut
  {NoStop}%
\bibitem [{Vul()}]{Vulcan}%
  \BibitemOpen
  \href@noop {} {\ }\bibinfo {note} {The Vulcan 10 Petawatt
  Project:\\http://www.clf.rl.ac.uk/New+Initiatives/The+Vulcan+10+Petawatt+Project/18577.aspx}\BibitemShut
  {NoStop}%
\bibitem [{ELI()}]{ELI}%
  \BibitemOpen
  \href@noop {} {\ }\bibinfo {note} {The Extreme Light Infrastructure (ELI)
  project:\\http://www.extreme-light-infrastructure.eu}\BibitemShut {NoStop}%
\bibitem [{\citenamefont {Di~Piazza}\ \emph {et~al.}(2012)\citenamefont
  {Di~Piazza}, \citenamefont {Muller}, \citenamefont {Hatsagortsyan},\ and\
  \citenamefont {Keitel}}]{DiPiazza:2011tq}%
  \BibitemOpen
  \bibfield  {author} {\bibinfo {author} {\bibfnamefont {A.}~\bibnamefont
  {Di~Piazza}}, \bibinfo {author} {\bibfnamefont {C.}~\bibnamefont {Muller}},
  \bibinfo {author} {\bibfnamefont {K.}~\bibnamefont {Hatsagortsyan}}\ and\
  \bibinfo {author} {\bibfnamefont {C.}~\bibnamefont {Keitel}},\ }\href
  {\doibase 10.1103/RevModPhys.84.1177} {\bibfield  {journal} {\bibinfo
  {journal} {Rev.Mod.Phys.}\ }\textbf {\bibinfo {volume} {84}},\ \bibinfo
  {pages} {1177} (\bibinfo {year} {2012})},\ \Eprint
  {http://arxiv.org/abs/1111.3886} {arXiv:1111.3886 [hep-ph]} \BibitemShut
  {NoStop}%
\bibitem [{\citenamefont {Bloch}\ and\ \citenamefont
  {Nordsieck}(1937)}]{Bloch:1937pw}%
  \BibitemOpen
  \bibfield  {author} {\bibinfo {author} {\bibfnamefont {F.}~\bibnamefont
  {Bloch}}\ and\ \bibinfo {author} {\bibfnamefont {A.}~\bibnamefont
  {Nordsieck}},\ }\href {\doibase 10.1103/PhysRev.52.54} {\bibfield  {journal}
  {\bibinfo  {journal} {Phys. Rev.}\ }\textbf {\bibinfo {volume} {52}},\
  \bibinfo {pages} {54} (\bibinfo {year} {1937})}\BibitemShut {NoStop}%
\bibitem [{\citenamefont {Kinoshita}(1962)}]{Kinoshita:1962ur}%
  \BibitemOpen
  \bibfield  {author} {\bibinfo {author} {\bibfnamefont {T.}~\bibnamefont
  {Kinoshita}},\ }\href@noop {} {\bibfield  {journal} {\bibinfo  {journal}
  {J.Math.Phys.}\ }\textbf {\bibinfo {volume} {3}},\ \bibinfo {pages} {650}
  (\bibinfo {year} {1962})}\BibitemShut {NoStop}%
\bibitem [{\citenamefont {Lee}\ and\ \citenamefont
  {Nauenberg}(1964)}]{Lee:1964is}%
  \BibitemOpen
  \bibfield  {author} {\bibinfo {author} {\bibfnamefont {T.~D.}\ \bibnamefont
  {Lee}}\ and\ \bibinfo {author} {\bibfnamefont {M.}~\bibnamefont
  {Nauenberg}},\ }\href {\doibase 10.1103/PhysRev.133.B1549} {\bibfield
  {journal} {\bibinfo  {journal} {Phys. Rev.}\ }\textbf {\bibinfo {volume}
  {133}},\ \bibinfo {pages} {B1549} (\bibinfo {year} {1964})}\BibitemShut
  {NoStop}%
\bibitem [{\citenamefont {Brock}\ \emph {et~al.}(1995)\citenamefont {Brock}
  \emph {et~al.}}]{Brock:1993sz}%
  \BibitemOpen
  \bibfield  {author} {\bibinfo {author} {\bibfnamefont {R.}~\bibnamefont
  {Brock}} \emph {et~al.} (\bibinfo {collaboration} {CTEQ Collaboration}),\
  }\href {\doibase 10.1103/RevModPhys.67.157} {\bibfield  {journal} {\bibinfo
  {journal} {Rev. Mod. Phys.}\ }\textbf {\bibinfo {volume} {67}},\ \bibinfo
  {pages} {157} (\bibinfo {year} {1995})}\BibitemShut {NoStop}%
\bibitem [{\citenamefont {Lavelle}\ and\ \citenamefont
  {McMullan}(2006)}]{Lavelle:2005bt}%
  \BibitemOpen
  \bibfield  {author} {\bibinfo {author} {\bibfnamefont {M.}~\bibnamefont
  {Lavelle}}\ and\ \bibinfo {author} {\bibfnamefont {D.}~\bibnamefont
  {McMullan}},\ }\href@noop {} {\bibfield  {journal} {\bibinfo  {journal}
  {JHEP}\ }\textbf {\bibinfo {volume} {03}},\ \bibinfo {pages} {026} (\bibinfo
  {year} {2006})},\ \Eprint {http://arxiv.org/abs/hep-ph/0511314}
  {arXiv:hep-ph/0511314} \BibitemShut {NoStop}%
\bibitem [{\citenamefont {Dinu}\ \emph {et~al.}(2012)\citenamefont {Dinu},
  \citenamefont {Heinzl},\ and\ \citenamefont {Ilderton}}]{Dinu:2012tj}%
  \BibitemOpen
  \bibfield  {author} {\bibinfo {author} {\bibfnamefont {V.}~\bibnamefont
  {Dinu}}, \bibinfo {author} {\bibfnamefont {T.}~\bibnamefont {Heinzl}}\ and\
  \bibinfo {author} {\bibfnamefont {A.}~\bibnamefont {Ilderton}},\ }\href
  {\doibase 10.1103/PhysRevD.86.085037} {\bibfield  {journal} {\bibinfo
  {journal} {Phys.Rev.}\ }\textbf {\bibinfo {volume} {D86}},\ \bibinfo {pages}
  {085037} (\bibinfo {year} {2012})},\ \Eprint {http://arxiv.org/abs/1206.3957}
  {arXiv:1206.3957 [hep-ph]} \BibitemShut {NoStop}%
\bibitem [{\citenamefont {Sengupta}(1952)}]{Sengupta1952}%
  \BibitemOpen
  \bibfield  {author} {\bibinfo {author} {\bibfnamefont {N.~D.}\ \bibnamefont
  {Sengupta}},\ }\href@noop {} {\bibfield  {journal} {\bibinfo  {journal}
  {Bull. Math. Soc. (Calcutta)}\ }\textbf {\bibinfo {volume} {44}},\ \bibinfo
  {pages} {175} (\bibinfo {year} {1952})}\BibitemShut {NoStop}%
\bibitem [{\citenamefont {Kibble}(1965)}]{Kibble:1965zza}%
  \BibitemOpen
  \bibfield  {author} {\bibinfo {author} {\bibfnamefont {T.}~\bibnamefont
  {Kibble}},\ }\href {\doibase 10.1103/PhysRev.138.B740} {\bibfield  {journal}
  {\bibinfo  {journal} {Phys.Rev.}\ }\textbf {\bibinfo {volume} {138}},\
  \bibinfo {pages} {B740} (\bibinfo {year} {1965})}\BibitemShut {NoStop}%
\bibitem [{\citenamefont {Reiss}\ and\ \citenamefont
  {Eberly}(1966)}]{Reiss:1966A}%
  \BibitemOpen
  \bibfield  {author} {\bibinfo {author} {\bibfnamefont {H.~R.}\ \bibnamefont
  {Reiss}}\ and\ \bibinfo {author} {\bibfnamefont {J.~H.}\ \bibnamefont
  {Eberly}},\ }\href {\doibase 10.1103/PhysRev.151.1058} {\bibfield  {journal}
  {\bibinfo  {journal} {Phys. Rev.}\ }\textbf {\bibinfo {volume} {151}},\
  \bibinfo {pages} {1058} (\bibinfo {year} {1966})}\BibitemShut {NoStop}%
\bibitem [{\citenamefont {Eberly}\ and\ \citenamefont
  {Reiss}(1966)}]{Eberly:1966b}%
  \BibitemOpen
  \bibfield  {author} {\bibinfo {author} {\bibfnamefont {J.~H.}\ \bibnamefont
  {Eberly}}\ and\ \bibinfo {author} {\bibfnamefont {H.~R.}\ \bibnamefont
  {Reiss}},\ }\href {\doibase 10.1103/PhysRev.145.1035} {\bibfield  {journal}
  {\bibinfo  {journal} {Phys. Rev.}\ }\textbf {\bibinfo {volume} {145}},\
  \bibinfo {pages} {1035} (\bibinfo {year} {1966})}\BibitemShut {NoStop}%
\bibitem [{\citenamefont {Ritus}(1972)}]{Ritus:1972ky}%
  \BibitemOpen
  \bibfield  {author} {\bibinfo {author} {\bibfnamefont {V.~I.}\ \bibnamefont
  {Ritus}},\ }\href {\doibase 10.1016/0003-4916(72)90191-1} {\bibfield
  {journal} {\bibinfo  {journal} {Annals Phys.}\ }\textbf {\bibinfo {volume}
  {69}},\ \bibinfo {pages} {555} (\bibinfo {year} {1972})}\BibitemShut
  {NoStop}%
\bibitem [{\citenamefont {Reiss}(2009)}]{Reiss:2009}%
  \BibitemOpen
  \bibfield  {author} {\bibinfo {author} {\bibfnamefont {H.~R.}\ \bibnamefont
  {Reiss}},\ }\href@noop {} {\bibfield  {journal} {\bibinfo  {journal} {Eur.
  Phys. J. D}\ }\textbf {\bibinfo {volume} {55}},\ \bibinfo {pages} {365}
  (\bibinfo {year} {2009})}\BibitemShut {NoStop}%
\bibitem [{\citenamefont {Harvey~C.}\ and\ \citenamefont
  {M.}(2012)}]{Harvey:2012ie}%
  \BibitemOpen
  \bibfield  {author} {\bibinfo {author} {\bibfnamefont {C.}\ \bibnamefont
  {Harvey}, \bibfnamefont {T.}\ \bibfnamefont {Heinzl}}, \bibfnamefont {A.}\ \bibfnamefont {Ilderton}\  and\ \bibfnamefont {M.}\ \bibinfo {author}
  {\bibfnamefont {Marklund}},\ }\href {\doibase
  10.1103/PhysRevLett.109.100402} {\bibfield  {journal} {\bibinfo  {journal}
  {Phys.Rev.Lett.}\ }\textbf {\bibinfo {volume} {109}},\ \bibinfo {pages}
  {100402} (\bibinfo {year} {2012})},\ \Eprint {http://arxiv.org/abs/1203.6077}
  {arXiv:1203.6077 [hep-ph]} \BibitemShut {NoStop}%
\bibitem [{\citenamefont {Boca}(2011)}]{Boca:2011m}%
  \BibitemOpen
  \bibfield  {author} {\bibinfo {author} {\bibfnamefont {M.}~\bibnamefont
  {Boca}},\ }\href {http://stacks.iop.org/1751-8121/44/i=44/a=445303}
  {\bibfield  {journal} {\bibinfo  {journal} {J. Phys. A: Math. Theor.}\
  }\textbf {\bibinfo {volume} {44}},\ \bibinfo {pages} {445303} (\bibinfo
  {year} {2011})}\BibitemShut {NoStop}%
\bibitem [{\citenamefont {Neville}\ and\ \citenamefont
  {Rohrlich}(1971)}]{Neville:1971uc}%
  \BibitemOpen
  \bibfield  {author} {\bibinfo {author} {\bibfnamefont {R.}~\bibnamefont
  {Neville}}\ and\ \bibinfo {author} {\bibfnamefont {F.}~\bibnamefont
  {Rohrlich}},\ }\href {\doibase 10.1103/PhysRevD.3.1692} {\bibfield  {journal}
  {\bibinfo  {journal} {Phys.Rev.}\ }\textbf {\bibinfo {volume} {D3}},\
  \bibinfo {pages} {1692} (\bibinfo {year} {1971})}\BibitemShut {NoStop}%
\bibitem [{\citenamefont {Ilderton}\ and\ \citenamefont
  {Torgrimsson}(2012)}]{Ilderton:2012qe}%
  \BibitemOpen
  \bibfield  {author} {\bibinfo {author} {\bibfnamefont {A.}~\bibnamefont
  {Ilderton}}\ and\ \bibinfo {author} {\bibfnamefont {G.}~\bibnamefont
  {Torgrimsson}},\ }\href@noop {} {\  (\bibinfo {year} {2012})},\ \Eprint
  {http://arxiv.org/abs/1210.6840} {arXiv:1210.6840 [hep-th]} \BibitemShut
  {NoStop}%
\bibitem [{\citenamefont {Krainov}\ \emph {et~al.}(1997)\citenamefont
  {Krainov}, \citenamefont {Reiss},\ and\ \citenamefont
  {Smirnov}}]{Krainov:book}%
  \BibitemOpen
  \bibfield  {author} {\bibinfo {author} {\bibfnamefont {V.}~\bibnamefont
  {Krainov}}, \bibinfo {author} {\bibfnamefont {H.}~\bibnamefont {Reiss}}\
  and\ \bibinfo {author} {\bibfnamefont {M.}~\bibnamefont {Smirnov}},\
  }\href@noop {} {\emph {\bibinfo {title} {Radiative Processes in Atomic
  Physics}}}\ (\bibinfo  {publisher} {Wiley},\ \bibinfo {year}
  {1997})\BibitemShut {NoStop}%
\bibitem [{\citenamefont {Raddadi}(2013)}]{Merfat}%
  \BibitemOpen
  \bibfield  {author} {\bibinfo {author} {\bibfnamefont {M.}~\bibnamefont
  {Raddadi}},\ }\href@noop {} {Ph.D. thesis},\ \bibinfo  {school} {Plymouth University} (\bibinfo {year} {2013})\BibitemShut {NoStop}%
\end{thebibliography}

%

\end{document}